\documentstyle[12pt]{article}

\newcommand{\be}{\begin{equation}}
\newcommand{\ee}{\end{equation}}
\newcommand{\bea}{\begin{eqnarray}}
\newcommand{\eea}{\end{eqnarray}}

\begin{document}
\begin{titlepage}

\begin{flushright}
{\today}
\end{flushright}
\vspace{1in}

\begin{center}
\Large
{\bf Analyticity Properties  of Scattering
Amplitude  in Theories with Compactified Space Dimensions     }
\end{center}

\vspace{.2in}

\normalsize

\begin{center}
{ Jnanadeva Maharana \footnote{Adjunct Professor, NISER, Bhubaneswar}  \\
E-mail maharana$@$iopb.res.in} 
\end{center}

\normalsize

\begin{center}
 {\em Institute of Physics \\
Bhubaneswar - 751005, India  \\
   and \\
   Max-Planck Institute for Gravitational Physics, Albert Einstein Institute, Golm, Germany  }

\end{center}

\vspace{.2in}

\baselineskip=24pt

\begin{abstract}
We consider a massive, neutral,  scalar field theory  of mass $m_0$ in  a five dimensional flat 
spacetime. Subsequently, one spatial dimension is compactified on a circle, $S^1$, of
radius $R$. The resulting theory  is defined in the manifold, $R^{3,1}\otimes S^1$.  The mass spectrum is a state  of  lowest mass, $m_0$, and
a tower of massive Kaluza-Klein states.  The analyticity property of the elastic scattering amplitude is investigated 
in the  Lehmann-Symanzik-Zimmermann (LSZ) formulation of this  theory. In the context of
nonrelativistic potential scattering, for the $R^3\otimes S^1$ spatial geometry, it was shown that the  forward scattering amplitude does not satisfy
analyticity properties in some cases for a class of potentials. If the same result is valid in relativistic quantum field theory then the 
consequences will be far reaching.
   We show that 
the forward elastic scattering amplitude 
of the  theory,  in the LSZ axiomatic approach,  satisfies forward dispersion relations. The importance of the unitarity constraint on the 
S-matrix, is exhibited in displaying
the properties of the absorptive part of the  amplitude.

\end{abstract}

\vspace{.5in}

\end{titlepage}


\noindent {\bf 1. Introduction}

\bigskip
\noindent
The analyticity property of scattering amplitude is a cardinal attribute and this  has been derived  
in the frameworks of general relativistic quantum 
field theories. The scattering amplitude, $F(s,t)$,  is an
analytic function of  the center of mass energy squared, $s$,   for
 momentum transfer squared, $t$. The dispersion relations in $s$ have been proved
when   $t$ is  within the Lehmann-Martin ellipse. This result has been   derived
from the axiomatic    approach of
Lehmann-Symanzik-Zimmenmann (LSZ) \cite{lsz} and  in the general frameworks of axiomatic formulation of field theories
\cite{book1,book2,book3, fr1,lehm1,sommer,eden,wight,jost,streat,kl,ss,bogo}.
The underlying structure of such theories are locality, mircocausality, Lorentz invariance to mention a few.
 It is generally accepted that if the dispersion relations are violated one might
question the validity  of the axioms of general field theories since these are the pillars on which the structure of  such field theories  rest. 
Moreover, there are
host of rigorous results, derived in the form of theorems, which  have been  tested against experimental
data. One of the most important result is the celebrated Froissart-Martin bound \cite{fr,andre} that the total cross section
is bounded from above:
$\sigma_t\le {{4\pi}\over{t_0}}(log s)^2$ where $t_0$ is determined from the first principles for a given scattering process.
 The high energy data for hadronic total cross sections respect 
 this bound over a wide range of energies accessible by  accelerators.\\
It is well known that the scattering amplitude in nonrelativistic potential scattering
enjoys certain analyticity properties in energy $k$ for a large class of potentials. This topic
has been studied quite thoroughly in the past \cite{gw,khuri1,wong}. One intriguing point is, in contrast
to relativistic quantum field theories (QFT), that the analyticity of the amplitudes in potential
scattering is not so intimately related to a principle like microcausality as is the case
with QFT. Moreover, the principle of microcausality has its root in the existence of the limiting
velocity in the  special theory of relativity i.e. velocity of light, $c$,  is the
limiting velocity.  Moreover, there is an intimate relationship between microcausality, as postulated in quantum field theories, 
and the analyticity properties of scattering amplitudes.
On the other hand, when we consider nonrelativistic quantum mechanical scattering in potential models the theory is  
invariant under the Galilean transformations. If we encounter a situation, in potential
scattering, where the amplitude fails to satisfy analyticity in the momentum, $k$, it is
not so much a matter of serious concern as would be the case for a relativistic QFT, especially the 
one which sastifies LSZ or Wightman axioms.\\
It is now generally accepted that theories which exist in higher spacetime dimensions, $\hat D>4$,
might  play important roles in our attempt to unify fundamental
interactions. We mention, in this context,  supersymmetric theories, supergravity theories and the string theories which have been
investigated intensively over past several decades.  It is generally postulated in the context of 
 such higher dimensional theories  that some of the spatial dimensions
 be compactified so that one eventually constructs an effective four dimensional theory of fundamental interactions to describe and 
 understand physics in the domain of present accessible energies.
It is proposed, in certain scenarios, that signatures of the extra spatial dimensions might be
observed in current high energy colliders \cite{anto,luest}.  Consequently, there has been a lot of
phenomenological studies to investigate and build models for possible experimental
observations of the decompactified dimensions at the present high energy accelerator such as LHC.
Indeed,  the scale of the extra compact dimensions is extracted from the  LHC experiments and it puts the 
compactification scale to be higher  than $600$ GeV. \\
Khuri \cite{khuri2}, motivated by the large compactification scenario,
had  set out to study analyticity properties of the scattering amplitude in potential scattering where
a spatial dimension is compactified on a circle; the so called $S^1$ compactification.
He discovered that the amplitude does not always satisfy the analyticity  properties. On the other hand the analyticity properties of the amplitude,
in the context of potential scattering, were investigated (for $d=3$) with noncompact spatial coordinates \cite{gw,khuri1,wong}
 i.e. there was no $S^1$ compactification. It was shown that 
the amplitude satisfied the dispersion relations.
 In fact it was shown by Khuri \cite{khuri2}, through
counter examples,  within the framework of perturbation theory, how the analyticity of the forward scattering amplitude
breaks down in the presence of $S^1$ compactification for a class of nonrelativistic potential models under certain circumstances
as we shall describe later. \\
The purpose of this investigation is to study the analyticity properties of  the scattering amplitude
in a field theory with an $S^1$ compactified spatial dimension. It is an analog of the Kaluza-Klein (KK) 
compactification although gravitational interaction is not incorporated. Moreover, after compactification,
we retain the entire tower of KK states. For sake of simplicity and to bring out the essential features,
we consider a single, neutral, massive scalar field  of mass $m_0$ in $D=5$. On compactification to $D=4$, not only
we have a massive scalar field with  mass $m_0$  as  of the 5-dimensional theory but also
we have tower of massive scalars as a consequence of compactification. \\
Recently, we have studied analyticity properties and the high energy behavior of  the  four point function
for a massive, neutral scalar field in higher dimensions,  $D>4$ \cite{jmjmp1}. It was shown, in the LSZ formalism,  that
the scattering amplitude has desire attributes in the following sense: (i)  We proved the generalization of the
Jost-Lehmann-Dyson theorem for the causal function and retarded function \cite{jl,dyson} for the $D>4$  case \cite{jmplb}. (ii) Subsequently, we
  showed the existence of the
Lehmann-Martin ellipse for such a theory. (iii) Thus a dispersion relation can be written  for the amplitude in $s$ for fixed $t$
when the momentum transfer squared lies inside Lehmann-Martin ellipse \cite{leh2,martin1}. (iv) The analog of Martin's theorem can be
derived in the sense that the scattering amplitude is analytic the product domain $D_s\otimes D_t$ where $D_s$ is the 
cut $s$-plane and $D_t$ is a domain in the $t$-plane such that the scattering amplitude is analytic inside a disk, $|t|<\tilde R$, $\tilde R$ is radius of the
disk and it is  independent of
$s$.  Thus the partial wave expansion converges inside this bigger domain. (v) We also derived the analog of Jin-Martin \cite{jml} upper bound on the scattering amplitude which states
that the fixed $t$ dispersion relation in $s$ does not require more than two subtractions. (vi) Therefore, a generalized
Froissart-Martin bound was be proved.\\
Our principal goal, in this investigation, is to examine the analyticity property of the 
four point function derived from the compactified theory in $D=4$, which originated 
from $D=5$ field theory, through the $S^1$ compactification. In other words whether the scattering amplitude 
 possesses  the desired analyticity properties as was derived for the higher dimensional
(uncompactified) theory. We shall specifically, investigate analyticity properties of the amplitude for scattering
of KK states in the forward direction in view of Khuri's result for scattering of such states in potential scattering
with  a spatial $S^1$ compactification. We  argue in sequel and prove that the forward scattering amplitude
satisfies desired analyticity property in $s$ and hence we are able to write down the dispersion relations.\\
 We work within the LSZ formalism for the effective four dimensional theory
obtained from the $D=5$ theory after $S^1$ compactification. The main conclusion is that, under certain assumptions,
the elastic forward scattering amplitude for the scattering of Kaluza-Klein states satisfies dispersion relation when a five 
dimensional massive theory is described on the manifold $R^{3,1}\otimes S^1$.\\
The paper is organized as follows. The next section (Section 2) is devoted to a brief review of the contents of Khuri's
paper \cite{khuri2}. The third section contains  the essentials of LSZ axiomatic approach to field theory defined
over a five dimensional spacetime. We define the retarded product of field operators (R-product) and elucidate their  properties.
 We introduce  kinematical variables which are
used in this paper. We also introduce notations and conventions followed in this article. The next section 
(Section 4),  deals with 
 the definitions, conventions and kinematics. Then we consider scattering of the 
 zero modes (the fields which carry lowest mass) of the 4-dimensional theory.  This problem is
similar to scattering of equal mass scalars in  a four dimensional theory. However, there are some subtleties since we
have to account for the  complete set of physical intermediate states (including entire KK tower) in certain spectral representations.
In   section 4.2.2 deals with elastic scattering of the lowest mass state with an excited  KK state. This is the case of unequal mass scattering
where one state carries KK charge (KK momentum) and the other state has no KK momentum. The fifth  section is
devoted to elastic scattering of two states carrying KK charges. Therefore, some caution is to be exercised in  deriving analyticity property of the amplitude.
We assume throughout this investigation that: (i) all particles are stable. (ii) There are no bound states. (iii) The KK charge (the discrete momentum
along the compact $S^1$ direction)  is conserved. 
Our conclusion is that the forward scattering amplitude possesses nice analyticity properties. Thus if we accept the
axioms of LSZ and adopt the standard procedures to investigate analyticity properties of the forward scattering
amplitudes we arrive at the same result as is known for  field theories defined in flat Minkowski spacetime and we show   that   the forward scattering
amplitude satisfies the dispersion relation. The sixth section summarizes our results and contains discussion. 
\newpage

\noindent {\bf 2. Non-relativistic Potential Scattering for $R^3\otimes S^1$ Geometry} 

\bigskip

\noindent  Khuri \cite{khuri2}  envisaged scattering of a particle in a space with $R^3\otimes S^1$ topology. We  provide a brief account
of his work and incorporate  his important conclusions. We refer the original paper to the  interested reader. The notations of \cite{khuri2} 
will be followed. 
The compactified coordinate is a circle of radius $R$ and it is assumed the radius is {\it small} i.e. ${1\over R}>> 1$ where dimensionless
units were used. The potential, $V(r,  \Phi)$, is such that is periodic in the angular coordinate, $ \Phi$, of $S^1$; $\bf r\in R^3$ and
$r=|{\bf r}|$. The potential $V(r,\Phi)$ belongs to a broad class such that for large $r$  these class of
potentials fall off like  $e^{-\mu r}/r$ as $r\rightarrow \infty$; $\mu>0$, carrying dimension of inverse length. 
 Moreover,  $V(r,\Phi)=V(r,\Phi+2\pi)$. The scattering amplitude depends on three variables - the momentum
of the particle, $k$, the scattering angle $\theta$, and an integer $n$ which appears due to the periodicity of the $\Phi$-coordinate.
Thus forward scattering amplitude is denoted by $T_{nn}(K)$, where $K^2=k^2+{{n^2}\over{R^2}}$.
The starting point is the Schr\" odinger equation
\bea
\label{kh1}
\bigg[{ \nabla}^2+{{1\over R^2}}{{\partial}^2\over{\partial\bf\Phi}^2} +K^2-V(r,\Phi )\bigg]\Psi({\bf r},\Phi)=0
\eea
The free plane wave solutions are 
\bea
\label{kh2}
\Psi_0({\bf x},\Phi})={{1\over{(2\pi)^2}}e^{i{\bf k}.{\bf x}}e^{i n\Phi}
\eea
and $n\in {\bf Z}$.
The total energy is defined to be
\bea
\label{kh3}
{\bf K}^2=k^2+{{n^2}\over{R^2}}
\eea
The free Green's function (in the presence of a compact coordinate) assumes the following form
\bea
\label{kh4}
G_0({\bf K};{\bf x},\Phi:{\bf x'},\Phi ')=-{{1\over{(2\pi)^4}}}\sum_{n=-\infty}^{n=+\infty}\int d^3p{{e^{i{\bf p}.({\bf x}-{\bf x'})}e^{in.(\Phi-\Phi')}}
\over{[p^2+{{n^2}\over{R^2}}-{\bf K}^2-i\epsilon]}} 
\eea
The free Green's function satisfies the free Schr\" odinger equation
\bea
\label{kh5}
\bigg[{ \nabla}^2+{{1\over R^2}}{{\partial}^2\over{\partial\bf\Phi}^2} +K^2\bigg] G_0({\bf K};{\bf x},\Phi:{\bf x'},\Phi ')=\delta^3({\bf x}-{\bf x'})\delta(\Phi -\Phi)
\eea
The $d^3p$ integration can be performed in the expression (\ref{kh4}) leading to
\bea
\label{kh6}
G_0({\bf K};{\bf x} -{\bf x'};\Phi-\Phi ')=-{{1\over{(8\pi^2)}}}\sum_{n=-\infty}^{n=+\infty}{{ e^{i{\sqrt{K^2-{(n^2/{R^2})}}|{\bf x}-{\bf x'}|}}}\over{|{\bf x}-{\bf x'}|}}
e^{in(\Phi-\Phi')}
\eea
Khuri introduced the prescription that ${\sqrt{K^2-{n^2/{R^2}}}}$ is defined in such a way that when  $n^2/{R^2}>K^2$
\bea
\label{kh7}
i{\sqrt{K^2-n^2/{R^2}}}\rightarrow -{\sqrt{n^2/{R^2}-K^2}},~~n^2>K^2R^2
\eea
Note that the series expansion for 
$G_0({\bf K};{\bf x} -{\bf x'};\Phi-\Phi ')$ as expressed in (\ref{kh6}) is strongly damped for large enough $|n|$. A careful analysis, as was carried
out in ref. \cite{khuri2},   shows that the Green's function is well defined and bounded, except for $|{\bf x}-{\bf x'}|\rightarrow 0$, similar 
to the properties of Green's functions in potential scattering for a fixed $K^2$.  Khuri \cite{khuri2}expressed 
 the scattering integral equation for the potential $V(r,\Phi)$ as
\bea
\label{kh8}
\Psi_{k,n}({\bf x},\Phi)=e^{i{\bf k}.{\bf x}}e^{in\Phi}+\int_0^{2\pi} d\Phi'\int d^3{\bf x'}G_0({\bf K};|{\bf x} -{\bf x'}|;|\Phi-\Phi '|)V({\bf x'},\Phi')\Psi_{k,n}({\bf x'},\Phi')
\eea
The expression for the scattering amplitude is extracted from the large $|{\bf x}|$ limit when one looks at the asymptotic behavior of 
the wave function, 
\bea
\label{kh9}
\Psi_{{\bf k},n}\rightarrow e^{{\bf k}.{\bf x}}e^{in\Phi}+ \sum_{m=-[KR]}^{+[KR]}T({\bf k'},m:{\bf k},n){{e^{ik'_{mn}|{\bf x}|}\over{|{\bf x}|}}}e^{im\Phi}
\eea
where $[KR]$ is the largest integer less than $KR$ and 
\bea
\label{kh10}
k'_{mn}={\sqrt{k^2+{{n^2}\over{R^2}}-{{m^2}\over{R^2}}}}
\eea
He also identifies a conservation rule:
 $K^2=k^2+(n^2/{R^2})=k'^2+(m^2/{R^2})$. Thus it is argued that  that the scattered wave has only $(2[KR]+1)$ components and those states with
$(m^2/(R^2)>k^2+(n^2/{R^2})$ are exponentially damped for large $|{\bf x}|$ and consequently these do not appear in the scattered wave (see eq. (\ref{kh7})).
Now the scattering amplitude is extracted from equations (\ref{kh8}) and (\ref{kh9}) to be
\bea
\label{kh11}
T({\bf k'},n';{\bf k},n)=-{{1\over{8\pi^2}}}\int d^3{\bf x'}\int_0^{2\pi} d\Phi'e^{-i{\bf k'}.{\bf x'}}e^{-in'\Phi'}V({\bf x'},\Phi')\Psi_{{\bf k},n}({\bf x'},\Phi')
\eea
The condition,  $k'^2+n'^2/{R^2}=k^2+n^2/{R^2}$ is to be satisfied. Thus the scattering amplitude describes the process where incoming wave
$|{\bf k},n>$ is scattered to final state $|{\bf k'},n'>$.\\
Remark: Reader should pay attention to the expression for the discussion of scattering processes in relativistic QFT in sequel and note
the similarities and differences in subsequent sections.  \\
Formally, the amplitude assumes the following form for the full Green's function
\bea
\label{kh12}
T({\bf k'},n';{\bf k},n)-T_B=&&-{{1\over{8\pi^2}}}\int ....\int d^3{\bf x}d^3{\bf x'} d\Phi d\Phi'e^{-i({\bf k'}.{\bf x'}+n'\Phi'}
V({\bf x'},\Phi')\nonumber\\&&G({\bf K};{\bf x'},{\bf x};\Phi',\Phi) V({\bf x},\Phi)e^{i({\bf k}.{\bf x}+n\Phi)}
\eea
Here $T_B$ is the Born term.
\bea
\label{kh12a}
T_B=-{1\over{8\pi^2}}\int d^3x\int_0^{2\pi}e^{i({\bf k}-{\bf k}).{\bf x}}V(x,\Phi)e^{i(n-n')\Phi}
\eea
 Full Green's function satisfied an equation with the full Hamiltonian
\bea
\label{kh13}
\bigg[{ \nabla}^2+{{1\over R^2}}{{\partial}^2\over{\partial\bf\Phi}^2} +K^2-V({\bf x},\Phi) \bigg]G({\bf K};{\bf x},{\bf x'},\Phi,\Phi' )=\delta^3({\bf x}-{\bf x'})\delta(\Phi-\Phi')
\eea
This is the starting point of computing scattering amplitude perturbatively in potential scattering \cite{gw}. Khuri \cite{khuri2} proceeds to study the analyticity
properties of the amplitude and it is a parallel development similar to investigations done in the past. In the context of theory with a compact space
dimension,  he analyses an amplitude like $T_{nn}(K)$ to the second order in the Born approximation.\\
Khuri explicitly computes the second Born term $T^{(2)}$ for the forward amplitude, for the choice $n=1$. He has discovered that the analyticity of the forward amplitude
breaks down with a counter example; where $T_{nn}(k)$ does not satisfy dispersion relations for a class of Yukawa-type
potentials of the form
\bea
\label{kh14}
V(r,\Phi)=u_0(r)+2\sum_{m=1}^N u_m(r)cos({m\Phi})
\eea
  where $u_m(r)=\lambda_m{{e^{-\mu r}} \over{ r}}$. Khuri noted an important feature of his studies  that in the case when scattering theory was applied 
  perturbatively in $R^3$ space the resulting amplitude satisfied  analyticity properties for similar Yukawa-type
  potentials \cite{khuri1,wong}. Thus there has been concerns\footnote{Andre Martin brought the work of Khuri \cite{khuri2} to my attention and persuaded me to undertake this
  invetstigation.}  when non-analyticity was discovered in the non-relativistic quantum mechanics
  in the space $R^3\otimes S^1$ by Khuri.\\
  We shall describe the framework of our investigation in the next section. We remark in passing that the analyticity of scattering amplitude in nonrelativitic
  scattering is not such a  profound property as in the relativistic QFT although the analyticity in non-relativistic potential scattering has been investigated
  quite thoroughly in the past \cite{gw}. However, it is to be noted that in absence a limiting velocity (in the relativistic case velocity of light, $c$, profoundly
  influences the study of the analyticity of amplitudes) the microcausality is not enforced in nonrelatvistic processes. As we shall show (and as has been 
  emphasized in many classic books on Quantum Field Theories) there is, indeed,  a deep connection between microcausality and analyticity. When a spatial dimension is
   compactified on $S^1$, the coordinate on the circle is periodic;   we can define concept of microcausality. We shall keep this aspect in mind and 
    we shall undertake a systematic study of the analyticity of scattering amplitude in the sequel.  
 
\newpage

\noindent{\bf 3. Field Theory in Five Dimensional Spacetime}

\bigskip

\noindent Let us consider a neutral, scalar field theory with mass, $m_0$, in flat five dimensional Minkowski space $R^{4,1}$. 
It is assumed that the particle is stable and there are no bound states. The notation is  that the  spacetime coordinates
are denoted as $\hat x$ and all operators are denoted with a  
 {\it hat} when they are defined in the five dimensional  space where the spatial coordinates 
are noncompact.The  LSZ axioms are \cite{lsz}:\\
{\bf A1.} The states of the system are represented in  a
Hilbert space, ${\hat{ \cal H}}$. All the physical observables are self-adjoint
operators in the Hilbert space, ${\hat{\cal H}}$.\\
{\bf A2.} The theory is invariant under inhomogeneous Lorentz transformations.\\
{\bf A3.} The energy-momentum of the states are defined. It follows from the
requirements of  Lorentz  and translation invariance that
we can construct a representation of the
orthochronous  Lorentz group. The representation
corresponds to unitary operators, ${\hat U}({\hat a},{\hat \Lambda})$,  and the theory is
invariant 
under these transformations. Thus there are hermitian operators corresponding
to spacetime translations, denoted as ${\hat P}_{{\hat\mu}}$, with ${\hat \mu}=0,1,2,3,4$ which have following
properties:
\be
\bigg[{\hat P}_{\hat\mu}, {\hat P}_{\hat\nu} \bigg]=0
\ee
If ${{\hat{\cal F}}({\hat x})}$ is any Heisenberg operator then its commutator with ${\hat P}_{\hat\mu}$
is
\be
\bigg[{\hat P}_{\hat\mu}, {\hat{\cal F}}({\hat x}) \bigg]=i{\hat\partial}_{\hat\mu} {\hat{\cal F}}({\hat x})
\ee
It is assumed that the operator does not explicitly depend on spacetime 
coordinates.
  If one chooses a representation where the translation operators, ${\hat P}_{\hat\mu}$,
are diagonal and the basis vectors $|{\hat p},\hat\alpha>$  span the Hilbert space,
${\hat{\cal H}}$, such that
\be
{\hat P}_{\hat\mu}|{\hat p},{\hat\alpha}>={\hat p}_{\hat\mu}|{\hat p},\hat\alpha>
\ee
then we are in a position to make more precise statements: \\
${\bullet}$ Existence of the vacuum: there is a unique invariant vacuum state
$|0>$ which has the property
\be
{\hat U}({\hat a},{\hat\Lambda})|0>=|0>
\ee
The vacuum is unique and is Poincar\'e invariant.\\
${\bullet}$ The eigenvalue of ${\hat P}_{\hat\mu}$, ${\hat p}_{\hat\mu}$,  
is light-like, with ${\hat p}_0>0$.
We are concerned  only with  massive stated in this discussion. If we implement
infinitesimal Poincar\'e transformation on the vacuum state then
\be
{\hat P}_{\hat\mu}|0>=0,~~~ {\rm and}~~~ {\hat M}_{\hat{\mu}\hat\nu}|0>=0
\ee
from above postulates. Note that ${\hat M}_{{\hat\mu}\hat\nu}$ are the generators of Lorentz
transformations.\\
{\bf A4.} The locality of theory implies that a (bosonic) local operator 
at spacetime point
${\hat x}^{\hat\mu}$ commutes with another (bosonic) 
local operator at ${\hat x}'^{\hat\mu}$ when  their
separation is spacelike i.e. if $({\hat x}-{\hat x}')^2<0$. Our Minkowski metric convention
is as follows: the inner product of two 5-vectors is given by
${\hat x}.{\hat y}={\hat x}^0{\hat y}^0-{\hat x}^1{\hat y}^1-...-{\hat x}^{4}{\hat y}^{4}$.
Since we are dealing with a neutral scalar
field, for the field operator ${\hat\phi }({\hat x})$: ${{\hat\phi}({\hat x})}^{\dagger}={\hat\phi}({\hat x})$ i.e.
 ${\hat\phi} ({\hat x})$ is hermitian.
By definition it  transforms as a scalar under inhomogeneous Lorentz
transformations 
\be {\hat U}({\hat a},{\hat\Lambda}){\hat\phi}({\hat x}){\hat U}({\hat a},{\hat\Lambda})^{-1}={\hat\phi}({\hat\Lambda} {\hat x}+{\hat a})
\ee
The micro causality, for two local field operators,  is stated to be 
\be
\bigg[{\hat\phi}({\hat x}),{\hat\phi}({\hat x}') \bigg]=0,~~~~~for~~({\hat x}-{\hat x}')^2<0
\ee
It is well known that, in the LSZ formalism,  we are concerned with vacuum
expectation values of time ordered products of operators as well as
with the  the retarded product of fields. The requirements of the above listed axioms
lead to certain relationship, for example, between vacuum expectation values of
R-products of operators. These are termed as linear relations and the importance of
the above listed axioms is manifested through these relations. In contrast, unitarity
imposes nonlinear constraints on amplitude. For example, if we expand an amplitude
in partial waves, unitarity demands  certain positivity conditions to be satisfied by
the partial wave amplitudes. \\
We summarize below some of the important aspects of LSZ formalism as we utilize them
through out the present investigation.\\
(i) The asymptotic condition: According to LSZ the field theory accounts for the asymptotic observables.
These correspond to particles of definite mass, charge and spin etc. ${\hat \phi}^{in}({\hat x})$ represents
the free field in the remote past and a  Fock space is generated by the field operator. The physical observable can be
expressed in terms of these fields.\\
(ii)  ${\hat\phi}( \hat x)$ is the interacting field. LSZ  technique incorporates a prescription to relate the
interacting field, ${\hat\phi}( \hat x)$, with ${\hat\phi}^{in}({\hat x})$; consequently, the asymptotic fields are
defined with a suitable limiting procedure. Thus we introduce the notion of the adiabatic switching off
of the interaction. A cutoff adiabatic function is postulated such that this function controls the 
interactions. It is $\bf 1$ at finite interval of time and it has a smooth limit of passing to zero
as $|t| \rightarrow \infty$. It is argued that when adiabatic switching is removed we can define
the physical observables.\\
(iii) The fields  ${\hat \phi}^{in}({\hat x})$ and ${\hat\phi}( \hat x)$ are related as follows:
\be
\label{z}
{\hat x}_0\rightarrow -\infty~~~~{\hat\phi}({\hat x})\rightarrow {\hat Z}^{1/2}{\hat\phi}^{in}({\hat x})
\ee
By the first postulate, ${\hat\phi}^{in}({\hat x})$ creates free particle states. However,
in general ${\hat\phi}({\hat x})$ will create multi particle states besides the single
particle one since it is the interacting field. Moreover, $<1|{\hat\phi}^{in}({\hat x})|0>$ 
and
 $<1|{\hat\phi}({\hat x})|0>$ carry same functional dependence in $\hat x$.  If the factor 
of $\hat Z$ were not the scaling relation between the two
fields (\ref{z}), then canonical commutation relation for each of the 
two fields ( i.e. ${\hat\phi}^{in}({\hat x})$ and  ${\hat\phi}({\hat x})$)  will be the same.
Thus in the absence of $\hat Z$ the two theories will be identical. Moreover, the
postulate of asymptotic condition states that in the remote future
\be
{\hat x}_0\rightarrow \infty~~~~{\hat\phi}({\hat x})\rightarrow {\hat Z}^{1/2}{\hat\phi}^{out}({\hat x}).
\ee
We may as well construct a Fock space utilizing ${\hat\phi}({\hat x})^{out}$ as we could with    ${\hat\phi}({\hat x})^{in}$
Furthermore, the vacuum is unique for ${\hat\phi}^{in}$,  ${\hat\phi}^{out}$ and ${\hat\phi}({\hat x})$. The
normalizable single particle states are the same i.e.
${\hat\phi}^{in}|0>={\hat\phi}^{out}|0>$. We do not display ${\hat Z}$ from now on. If at all
any need arises,  ${\hat Z}$ can be introduced in the relevant expressions.\\
We define creation and annihilation operators for ${\hat\phi}^{in}$,
${\hat\phi}^{out}$. We recall that   ${\hat\phi}({\hat x})$  is not a free field. Wheheas the fields ${\hat\phi}^{in,out}({\hat x})$ satisfy the free field 
equations $[{\Box}_5+m^2_0]{\hat\phi}^{in,out}({\hat x})=0$; the interacting field satisfies an equation of motion
which is endowed with  a source current: $[{\Box}_5+m^2_0]{\hat\phi}({\hat x})]={\hat j}({\hat x})$.
 We  may use  the plane wave basis for simplicity in certain computations; however,
in a more formal approach, it is desirable to use wave packets.\\
The relevant vacuum expectation values of the products of operators in LSZ formalism are either the time ordered
products (the T-products) or the retarded products (the R-products). We shall mostly use the R-products and 
we use them extensively  throughout this investigation. It is defined as 
\bea
R~{\hat\phi}({\hat x}_0){\hat\phi}_1({\hat x}_1)...{\hat\phi}_n({\hat x}_n)=&&(-1)^n\sum_P\theta({\hat x}_{00}-{\hat x}_{10})
\theta({\hat x}_{10}-{\hat x}_{20})...\theta({\hat x}_{n-10}-{\hat x}_{n0})\nonumber\\&&
[[...[{\hat\phi}({\hat x}),{\hat\phi}_{i_1}({\hat x}_{i_1})],{\hat\phi}_{i_2}({\hat x}_{i_2})]..],{\hat\phi}_{i_n}({\hat x}_{i_n})]
\eea
note that $R{\hat\phi}({\hat x})={\hat\phi}({\hat x})$ and  P stands for all the permutations ${i_1,....i_n}$  of  $1,2...n$.
The R-product is hermitian for hermitian fields ${\hat\phi}_i({\hat x}_i)$ and
the product is symmetric under exchange of any fields
${\hat\phi}_1({\hat x}_1)...{\hat\phi}_n({\hat x}_n)$. Notice that the field ${\hat\phi}({\hat x})$ is kept where it is
located in  its position.
We list below some of the important properties of the $R$-product for future use \cite{fr1}:\\
(i) $R~{\hat\phi}({\hat x}_0){\hat\phi}_1({\hat x}_1)...{\hat\phi}_n({\hat x}_n) \ne 0$ only if
${\hat x}_{00}>~{\rm max}~\{{\hat x}_{10},..{\hat.x}_{n0} \}$.\\
(ii) Another important property of the R-product is that
\be
R~{\hat\phi}({\hat x}_0){\hat\phi}_1({\hat x}_1)...{\hat\phi}_n({\hat x}_n) = 0
\ee
whenever the time component ${\hat x}_{00}$, appearing in the argument of ${\hat\phi}({\hat x}_0)$ whose
position is held fix, is less than time component of any of the four vectors
$({\hat x}_1,...{\hat x}_n)$ appearing in the arguments of ${\hat\phi}({\hat x}_1)...{\hat\phi}({\hat x}_n)$.\\
(iii) We recall that under a Lorentz transformation ${\hat U}({\hat\Lambda},0)$
\be
{\hat\phi}({\hat x}_i)\rightarrow {\hat\phi}({\hat\Lambda} {\hat x}_i)={\hat U}({\hat\Lambda},0){\hat\phi}({\hat x}_i){\hat U}({\hat\Lambda},0)^{-1}
\ee
 Therefore,
\be
R~{\hat\phi}({\hat\Lambda}{\hat x}){\hat\phi}({\hat\Lambda}{\hat x}_i)...{\hat\phi}_n({\hat\Lambda} {\hat x}_n)={\hat U}({\hat\Lambda},0)
R~\phi(x)\phi_1(x_1)...\phi_n(x_n)U(\Lambda,0)^{-1}
\ee
And
\be
 {\hat\phi}_i({\hat x}_i)\rightarrow{\hat\phi}_i({\hat x}_i+{\hat a})=e^{i{\hat a}.{\hat P}}{\hat\phi}_i({\hat x}_i)e^{-i{\hat a}.{\hat P}}
\ee
 under spacetime translations. Consequently,
\be
R~{\hat\phi}( {\hat x}+{\hat a}){\hat\phi}({\hat x}_i+{\hat a})...{\hat\phi}_n({\hat x}_n+{\hat a})=
e^{i{\hat a}.{\hat P}}R~{\hat\phi}({\hat x}){\hat\phi}_1({\hat x}_1)...{\hat\phi}_n({\hat x}_n)e^{-i{\hat a}.{\hat P}}
\ee
 Therefore,   the vacuum expectation value of the R-product
dependents only on  difference between pair of coordinates: in other words it
depends on the
following set of coordinate differences: 
${\hat\xi}_1={\hat x}_1-{\hat x}_0,{\hat\xi}_2={\hat x}_2-{\hat x}_1...{\hat\xi}_n={\hat x}_{n-1} -{\hat x}_n$ as a consequence of
translational invariance. \\
(iv) The retarded property of R-function and the asymptotic conditions lead
 to the following relations.
\bea
[R~{\hat\phi}({\hat x}){\hat\phi}_1({\hat x}_1)...{\hat\phi}_n({\hat x}_n),{\hat\phi}^{in}_l({\hat y}_l)]=
i\int d^5{\hat y}'_l\Delta({\hat y}_l-{\hat y'_l})({\Box}_{{5\hat y}'}+{\hat m}_l^2)
R~{\hat\phi}({\hat x}){\hat\phi}_1({\hat x}_1)...{\hat\phi}_n({\hat x}_n){\hat\phi}_l({\hat y}'_l)
\eea
Note: here ${\hat m}_l$ stands for the mass of a field in five dimensions.
We may define 'in' and 'out' states in terms of the creation operators associated with 'in' and 'out' fields as follows
\be
\label{fock1}
|{\hat k}_1,{\hat k}_2,....{\hat k}_n~in>={\hat a}_{in}^{\dagger}({\hat{\bf k}}_1){\hat a}_{in}^{\dagger}({\hat{\bf k}}_2)...
{\hat a}_{in}^{\dagger}({\hat{\bf k}}_n)|0>
\ee
\be
\label{fock2}
|{\hat k}_1,{\hat k}_2,....{\hat k}_n~out>={\hat a}_{out}^{\dagger}({\hat{\bf k}}_1){\hat a}_{out}^{\dagger}({\hat{\bf k}}_2)...
{\hat a}_{out}^{\dagger}({\hat{\bf k}}_n)|0>
\ee
We can construct a complete set of states either starting from 'in'  field operators or the 'out' field operators and each complete set
will span the Hilbert space,  ${\hat{\cal H}}$. Therefore, a unitary operator will relate the two sets of states in this Hilbert
space. This is a heuristic way of introducing the concept of the $S$-matrix. We shall define $S$-matrix elements
through LSZ reduction technique in subsequent section.\\
We shall not distinguish between notations like ${\hat\phi}^{out,in}$ or ${\hat\phi}_{out,in}$ and therefore, there might be use
of the sloppy notation in this regard.\\
We record the following important remark {\it en passant}: The generic matrix element 
$<{\hat\alpha}|{\hat\phi}({\hat x}_1){\hat\phi}({\hat x}_2)...|{\hat\beta}>$
is not an ordinary function but a distribution. Thus it is to be always
understood as smeared with a Schwarz type test function $f\in {\cal S}$. The test
function is infinitely differentiable and it goes to zero along with all its
derivatives faster than any power of its argument. We shall formally derive expressions
for scattering amplitudes and the absorptive parts by employing the LSZ technique. It is to be understood that
these are generalized functions and such matrix elements are properly defined
with smeared out test functions.\\
We  obtain below the expression for the K\"allen-Lehmann representation for the
five dimensional theory. It will help us to transparently expose, as we shall recall in the next section, the consequences of
$S^1$ compactification. Let  us consider the vacuum expectation value (VEV) of the
commutator of two fields in the $D=5$ theory: $<0|[{\hat\phi}({\hat x}), {\hat\phi}({\hat y})]|0>$. We
introduce a complete set of states between product of the fields after opening up the commutator. Thus
we arrive at the following expression by adopting the standard arguments,
\bea
\label{KL}
  <0|[{\hat\phi}({\hat x}), {\hat\phi}({\hat y})]|0>=\sum_{\hat\alpha}\bigg(<0|{\hat\phi}(0){\hat\alpha}>e^{-i{\hat p}_{\hat\alpha}.({\hat x}-{\hat y})}
<{\hat\alpha}|{\hat\phi}(0)|0>-({\hat x}{\leftrightarrow}{\hat y}) \bigg)
\eea
Let us define
\be
{\hat\rho}({\hat q})=(2\pi)^4\sum_{\hat\alpha}\delta^5({\hat q}-{\hat p}_{\hat\alpha})|<0|{\hat\phi}(0)|{\hat\alpha}>|^2
\ee
Note that ${\hat\rho}({\hat q})$ is positive, and ${\hat\rho}=0$ when ${\hat q}$ is not in the light cone. It is also Lorentz
invariant. Thus we write
\be
{\hat\rho}({\hat q})={\hat\sigma}({\hat q}^2)\theta({\hat q}_0),~~{\hat\sigma}({\hat q}^2)=0,~~~if~~{\hat q}^2<0
\ee
This is a positive measure. We may separate the expression for the VEV of the commutator (\ref{KL}) into two parts:
the single particle state contribution and the rest. Moreover, we use the asymptotic state condition to arrive at
\bea
\label{KL1}
<0|[{\hat\phi}(\hat x}),{\hat\phi}({\hat y})]|0>=i{\hat Z}{\hat\Delta}({\hat x},{\hat y}; m_0)+i\int_{{\hat m}_1^2}^{\infty}d{\hat m}'^2{\hat{\Delta}
({\hat x},{\hat y}; {\hat m}')
\eea
where ${\hat\Delta}({\hat x},{\hat y};m_0)$ is the VEV of the free field commutator, $m_0$ is the mass of the scalar. ${\hat m}_1^2>{\hat M}^2$, the multiple
particle threshold.\\
We are in a position to  study several attributes of scattering amplitudes in the five dimensional theory such as proving existence of
the Lehmann-Martin ellipse, give a proof of fixed t dispersion relation to mention a few. However, these properties
have been derived in a general setting recently \cite{jmjmp1} for D-dimensional theories.
The purpose of incorporating the expression for the VEV of the commutator of two fields in the 5-dimensional theory
is to provide a prelude to the modification of similar expressions when we compactify the theory on $S^1$ as we shall
see in the next section.

\bigskip

\noindent{\bf 4. Compactification of the Scalar Field Theory: } ${\bf R^{4,1} \rightarrow  R^{3,1}\otimes S^1}$

\bigskip

\noindent We consider $S^1$ compactification of a spatial coordinate of the five dimensional theory. Let us decompose the
five dimensional spacetime coordinates, ${\hat x}^{\hat\mu}$,  as follows:
\be
 {\hat x}^{\hat\mu}=(x^{\mu}, y), \mu=0,1,2,3
  \ee
  where $x^{\mu}$ are the four dimensional Minkowski  space coordinates; $y$ is the spatial coordinate defined on $S^1$
  such that $y+2\pi R = y$, $R$ being the radius of compactification. We shall capture the essential features of the $S^1$
  compactification when a neutral scalar field ( in $D=5$) of mass $m_0$ is described in the geometry $R^{3,1}\otimes S^1$. 
  Let us consider as  a first step, some properties of the asymptotic fields  such as the { 'in' and 'out' field} ,  ${\hat\phi}^{in, out}({\hat x})$. The equation of motion is
  $[{\Box}_5+m_0^2]{\hat\phi}^{in,out}({\hat x})=0$. We expand the fields as follows 
  \be
  \label{kk1}
  {\hat\phi}^{in,out}({\hat x})={\hat\phi}^{in,out}(x,y)=\phi^{in,out}_0(x)+\sum_{n=-\infty}^{+n=\infty}\phi^{in,out}_n(x)e^{{{in y}\over{ R}}}
  \ee
  Note that $\phi^{in,out}_0(x)$ has no $y$-dependence and it is the so called {\it  zero mode}. The terms in rest of the series
  satisfy periodicity in $y$. We can decompose the five dimensional ${\Box}_5$ as sum of a 
   four dimensional $\Box$ and a ${{\partial}\over{\partial y^2}}$ term. The equation of motion is
  \be
  \label{kk2}
  [\Box - {{\partial}\over{\partial y^2}}+m_n^2]\phi^{in,out}_n(x,y)=0
  \ee
  where $\phi^{in,out}_n(x,y)=\phi_n^{in,out}e^{{{in y}\over{ R}}}$ and $n=0$ term has no $y$-dependence being $\phi_0(x)$.
  Here $m_n^2=m_o^2+{{n^2}\over{R^2}}$. Thus we have tower of massive states. The momentum associated
  in the $y$-direction is $q_n=n/R$ and is quantized  in the units of $1/R$ and it is an additive conserved quantum
  number. We term it as Kaluza-Klein (KK) charge although there is no gravitational interaction in the five dimensional theory;  we
   still call it KK reduction.  For the interacting field ${\hat\phi}({\hat x})$, we can adopt a similar mode expansion. 
   \be
   \label{kk2x}
   {\hat\phi}({\hat x})={\hat\phi}(x,y)=\phi_0(x)+\sum_{n=-\infty}^{n=+\infty}\phi_n(x)e^{{{iny}\over R}}
   \ee
   The equation of motion for the interaction fields in endowed with a source term. Thus source current would be expanded
   as is the expansion (\ref{kk2x}). Each field $\phi_n(x)$ will have a current, $J_n(x)$ associated with it and source
   current will be expanded as
   \be
   \label{kk2a}
   {\hat j}(x,y) =j_0(x)+\sum_{n=-\infty}^{n=+\infty}J_n(x)e^{i{ny/R}}
   \ee
   Note that the set of currents,  $\{J_n(x)  \}$, are the source currents associated with the tower of interacting fields
   $\{ \phi_n(x) \}$, $n\ne 0$. These fields carry the discrete KK charge, $n$. Therefore, $J_n(x)$ also carries the
   same KK charge. We should keep this aspect in mind when we consider matrix element of such currents between
   states. In future, we might not explicitly display the charge of the current; however, it becomes quite obvious
   in the context.\\ 
  The zero modes, $\phi^{in,out}_0$, create their  Fock spaces. Similarly, each of the fields $\phi^{in,out}_n(x)$ create  their Fock spaces as well.
  For example a state with spatial momentum, ${\bf p}$, energy, $p_0$ and discrete momentum $q_n$ (in $y$-direction) is created
  by
  \be
   \label{kk3}
   A^{\dagger, in}({\bf p},q_n)|0>=|p,q_n>_{in},~~p_0>0 
   \ee
   and similar orthogonality relation holds for  an {\it out} state.
  We may recall that in the five dimensional theory, we started with, there was only one neutral scalar field in the spectrum.
  As a consequence of the $S^1$ compactification, the resulting spectrum consists of a massive neutral scalar of mass $m_0^2$ and
  a tower of 'charged' massive field. Moreover, each level in this tower is characterized by a mass and a 'charge', $(m_n, q_n)$,
  respectively; the zero mode has $q_n=0$. Let us consider the Hilbert space of the compactified theory, keeping in mind the above remarks. \\
   {\it The Decomposition of the Hilbert space ${\hat{\cal H}}$:} The Hilbert space associated with the five dimensional theory is ${\hat {\cal H}}$. It is now decomposed as a direct sum of Hilbert spaces where each one is characterized by its $q_n$ quantum number  
   of the compactified theory 
  \be
  \label{kk4}
  {\hat{\cal H}}=\sum \oplus {\cal H}_n
  \ee
  Thus ${\cal H}_0$ is the Hilbert space which is spanned by states built from the creation operators $\{ a^{\dagger}({\bf k}) \}$ which acting
  on the vacuum create the complete set of states that span ${\cal  H}_0$. 
   A single particle state is  $a^{\dagger,in}({\bf k})|0>=|{\bf k}>_{in}$ and multiparticle states
  are created using the procedure out lines in (\ref{fock1}) and (\ref{fock2}) starting from {\it in} field.
   If we consider a field $\phi^{in}_n(x,y)$ with a charge $q_n$, we can create
  a Fock space through the set of creation operators $\{A^{\dagger,in}({\bf p}, q_n)\}$. Moreover,  two  state vectors with different 'charges'
  are orthogonal: for example 
  \be
  \label{kk5}
  <{\bf p}, q_{n'};in|{\bf p}',q_{n'};in>=\delta^3({\bf p}-{\bf p'})\delta_{n,n'}
   \ee
   We could as well create a Fock space utilizing {\it out} fields.\\ 
  {\it Remark: }We have stated earlier and repeat here that we assume that there are no bound states in the theory and all particles
  are stable. There exists a possibility that a particle with charge $2n$ and mass $m^2_{2n}=m_0^2+{{4n^2}\over{R^2}}$ could be a
  bound state of two particles of charge $n$ and masses $m_n$ each under certain circumstances. We have excluded such possibilities
  from the present investigation. \\
  Now we can adopt the LSZ formalism for the four dimensional spacetime with an extra compact dimension. If we keep in mind the steps
  introduced above, it is possible to envisage field operators $\phi^{in}_n(x)$ and $\phi^{out}_n(x)$ for each of the fields for a given KK charge, $n$.
  Therefore, each Hilbert space, ${\cal H}_n$ will be spanned by the state vectors  (say for 'in' states) created by operators $a^{\dagger, in}({\bf k})$, for $n=0$ and
  $A^{\dagger, in}({\bf p}, q_n)$, for $n\ne 0$.  Moreover, we are in a position to define corresponding set of  {\it interacting fields} $\{\phi_n(x) \}$ which
  will interpolate  into 'in' and 'out' fields in the asymptotic limits designated by their KK charges.\\
  {\it Remark}: Note that in (\ref{kk1}) sum over $\{n\}$ runs over positive and negative integers. If there is a parity symmetry, $y\rightarrow -y$, 
  under which the field is invariant we can reduce the sum to positive $n$ only. However, since $q_n$ is an additive discrete quantum number, 
  a state with $q_n>0$ could be designated as a particle and the corresponding state $q_n<0$ can be interpreted as its antiparticle. Thus
  a two particle state $|p,q_n>|p,-q_n>,~~q_n>0~and ~p_0>0$ is a particle antiparticle state, $q_n=0$.   For example,
  it could be two particle state of $\phi_0$ satisfying energy momentum conservation, especially if they appear as intermediate states. We shall
  keep this fact in mind for future references.     
   \\
   Let us momentarily return to the K\"allen-Lehmann representation (\ref{KL}) in the present context and utilize the expansion (\ref{kk2x}) in the 
   expression for the VEV of the commutator of two fields defined in $D=5$:  $<0|[{\hat\phi}({\hat x}),{\hat\phi}({\hat x}')]|0>$
   \bea
   \label{kk6}
  <0|[{\hat\phi}(x,y),{\hat\phi}(x',y')]|0>=<0|[\phi_0(x)+\sum_{-\infty}^{+\infty}\phi_n(x,y),~ \phi_0(x')+\sum_{-\infty}^{+\infty}\phi_l(x',y')]|0> 
   \eea
   The VEV of a commutator of two fields  given by the  spectral representation (\ref{KL}) will be decomposed into sum of several commutators
   whose VEV will appear:
   \bea
   \label{kk7}
   <0|[\phi_0(x), \phi_0(x')]|0>,~~<0|[\phi_n(x),\phi_{-n}(x')]|0>,...
   \eea
   Since the vacuum carries zero KK charge, $q_n=0$, the commutator of two fields (with $n\ne 0$) should give rise to zero-charge and  consequently, 
   only $\phi_n$ and $\phi_{-n}$ commutators will appear.
    Moreover, commutator of fields with different $q_n$ vanish since the operators act on states of different Hilbert spaces. Thus
   we already note the consequences of compactification. When we wish to evaluate the VEV and insert complete set of intermediate
   states in the product of two operators after opening up the commutators, we note that all states of the entire KK tower can appear
   as intermediate states as long as they respect all conservation laws. This will be an important feature in all our computations in
   what follows. 
   
   \bigskip

   \noindent{\bf{4.1 Conventions, Definitions and Kinematics}}

   \bigskip
   
   \noindent We have stated earlier that our goal is to study the analyticity property of the four point amplitude in the {\it foward} direction.
   So far we have laid down the requisite procedures for compactification and we have outlined the structure of the Hilbert space in the compactified
   theory. We defined {\it in} and {\it out} fields in each of the sectors. Thus we can apply the LSZ reduction technique to derive expressions for the
   scattering amplitudes keeping in mind the energy momentum conservation rules and conservation of the KK charge.\\
   We adopt the following notations: the field associate with the zero mode (earlier denoted as $\phi_0$) is denoted as $\phi$. If we consider
   scattering of four such particles for the process $a+b\rightarrow c+d$, all being $\phi$ fields we shall denote it as 
   $\phi_a+\phi_b\rightarrow \phi_c+\phi_d$. The four momenta of $\phi$ particles will be denoted as $k$, in the preceding reaction  it will
   be denoted as  $k_a+k_b\rightarrow k_c+k_d$.
    The creation and annihilation operators (say for the 'in' fields) are: $a^{\dagger,in}({\bf k})$ and $a^{in}({\bf k})$ respectively.
   Any field which belongs to the KK tower is denoted by $\chi_n$, $n$ being the KK charge it carries and we always denote four momentum of 
   a KK particle as $p_{\mu}$. {\it For  conveniences, we use the notation $q_n=n$ from now on which amounts to adopting the convention that $R=1$}.
   For any (say 'in') field belonging to KK-tower, creation and annihilation operators are denoted respectively as $A^{\dagger,in}({\bf p}, q_n)$ and $A^{in}({\bf p},q_n)$. Sometimes,
   we might not explicitly exhibit presence of the KK charge in a reaction; however, it will stated in a context whenever required.There will be three types of 
   scattering processes.\\
   (i) $q_n=0$ sector: the reaction involves only four $\phi$ fields: $\phi+\phi\rightarrow \phi+\phi$.\\
   (ii) Scattering of a $\phi$ field with a $\chi$ field such as $\phi+\chi_n\rightarrow \phi+\chi_n$. Since KK charge is conserved, by  assumption, the
   initial and final state particles are described by the  $\chi$ fields with the same charge.\\
   (iii) The scattering of four $\chi$ fields. They could be of two types: (a) Elastic scattering where $\chi_n+\chi_m\rightarrow\chi_n+\chi_m$.
   Here the initial particles carry KK charges $n$ and $m$ and final particles also carry same charges. (b)
   Inelastic scattering like $\chi_n+\chi_m\rightarrow \chi_{n'}+\chi_{m'}$. The total KK charge conservation implies $n+m=n'+m'$.\\
   Of course in all reactions, total energy momentum conservation is to be guaranteed.
\\
Let us consider a generic 4-body reaction
\be
\label{kk6}
{\tilde a}+{\tilde b}\rightarrow {\tilde c}+{\tilde d} 
\ee
The particles $({\tilde a},{\tilde b}, {\tilde c},{\tilde d} )$ (the corresponding fields being ${\tilde\phi}_a, {\tilde\phi}_b, {\tilde\phi}_c, {\tilde\phi}_d$)
respectively carrying momenta $ {\tilde p}_a, {\tilde p}_b, {\tilde p}_c, {\tilde p}_d$; these particles may
correspond to the KK zero modes (with KK momentum $q=0$) or particles might carry nonzero KK charge. We shall consider
{\it only elastic scatterings}. The Lorentz invariant Mandelstam variables are
\be
\label{kk7}
s=({\tilde p}_a+{\tilde p}_b)^2=({\tilde p}_c+{\tilde p}_d)^2,~t=({\tilde p}_a-{\tilde p}_d)^2=({\tilde p}_b-{\tilde p}_c)^2,~
u=({\tilde p}_a-{\tilde p}_c)^2=({\tilde p}_b-{\tilde p}_d)^2
\ee
and $\sum {\tilde p}^2_a+{\tilde p}^2_b+{\tilde p}^2_c+{\tilde p}^2_d=m_a^2+m_b^2+m_c^2+m_d^2$. We shall
maintain independent identities of the four particles which will facilitate the computation of the four point function
utilizing the LSZ reduction technique.  We list below some relevant (kinematic) variables which we need for our
future discussions:
\be
\label{kk8}
{\bf M}_a^2,~~{\bf M}_b^2,~~{\bf M}_c^2,~~{\bf M}_d^2
\ee
These correspond to lowest mass two or more particle states which carry the same quantum number as that of
particle $a$, $b$, $c$ and $d$ respectively. We also define six more variables to be
\be
\label{kk9}
({\bf M}_{ab}, {\bf M}_{cd}),~~({\bf M}_{ac}, {\bf M}_{bd}),~~({\bf M}_{ad}, {\bf M}_{bc})
\ee
The variable ${\bf M}_{ab}$ carries the same quantum number as $(a ~and~ b)$ and it corresponds to two or more particle
states. Similar definition holds for the other five variables introduced above. 
 We define two types of thresholds: (i) the physical threshold, $s_{phys}$, and $s_{thr}$. In absence of anomalous thresholds (and equal mass scattering)
 $s_{thr}=s_{phys}$ for a reaction to proceed in the $s$ channel. 
 Similarly, we may define $u_{phys}$ and $u_{thr}$ which will be useful when we discuss dispersion relations.  We assume from 
 now on that $s_{thr}=s_{phys}$ and $u_{thr}=u_{phys}$.
We shall outline how a four point function is obtained in LSZ approach.  Normally one starts with 
$|{\tilde p}_d,{\tilde p}_c~out>$ and $|{\tilde p}_b,{\tilde p}_a~in>$ and considers the matrix element $<{\tilde p}_d,{\tilde p}_c~out| {\tilde p}_b,{\tilde p}_a~in>$.
Then we subtract out the matrix element $<{\tilde p}_d,{\tilde p}_c~in|{\tilde p}_b,{\tilde p}_a~in>$ to define the S-matrix element.
\bea
\label{kk10}
<{\tilde p}_d,{\tilde p}_d~out|{\tilde p}_b,{\tilde p}_a~in>=&&\delta^3({\tilde{\bf p}}_d-{\tilde{\bf p}}_b)\delta^3({\tilde{\bf p}}_c-{\tilde{\bf p}}_a)
-{{i}\over{(2\pi)^3}} \int d^4x\int d^4x' \nonumber\\&&e^{-i({\tilde p}_a.x-{\tilde p}_cx')}K_xK_{x'}<{\tilde p}_d~out|R(x',x)|{\tilde p}_b~in>
\eea
where $K_x$ and $K_{x'}$ are the four dimensional Klein-Gordon operators and
\be
\label{kk11}
R(x,x')=-\theta (x_0-x_o')[{\tilde\phi}_a(x),{\tilde\phi}_c(x')]
\ee
We have reduced fields associated with $a$ and $c$ in (\ref{kk10}). 
In the next step we may reduce all the four fields and in such a reduction we shall get VEV of the R-product of four fields which will be operated
upon by four K-G operators. However, the latter form of  LSZ reduction (when all fields are reduced) 
is not generally utilized  when we want to investigate the analyticity property of the
amplitude in the present context.
 In particular our intent is  to write the forward dispersion relation. Thus we abandon the idea of reducing all the four fields in this article.\\
{\it Remark: } Note that on the right hand side of the requation (\ref{kk10}) the operators act on
 $R{\tilde\phi}_a(x){\tilde\phi}(x')_c$ and there is a $\theta$-function
in the definition of the R-product. Consequently, the action of $K_xK_{x'}$ on ${\tilde\phi}_a(x){\tilde\phi}_c(x')$ will produce a term 
$R{\tilde j}_a(x){\tilde j}_c(x')$. In addition the operation of the two K-G operators will give rise to $\delta$-functions and derivatives of $\delta$-functions
and some equal time commutators i.e. there will terms whose coefficients are $\delta (x_0-x_0')$. When we consider fourier transforms of 
the derivatives of these $\delta$-function derivative terms they will be transformed to momentum variables. However, the amplitude is a function of
Lorentz invariant quantities. Thus one will get only finite polynomials of such variables, as has been argued by Symanzik \cite{kurt}.
His arguments is that  in a local quantum field
theory only finite number of derivatives of $\delta$-functions can appear. Moreover, in addition, there are some 
equal time commutators and many of them vanish when we
invoke locality arguments. Therefore, we shall use the relation
\be
\label{kk12}
K_xK_{x'}R{\tilde\phi}(x),{\tilde\phi}_c(x')=R{\tilde j}_a(x){\tilde j}_c(x')
\ee
 keeping in mind that there are derivatives of $\delta$-functions and some equal time commutation relations which might be present on the right hand side
 of the above equation.
 Moreover, since the derivative terms give rise to polynomials in Lorentz invariant variables, the analyticity properties of the amplitude
 are not affected due to the presence of such terms. This will be understood whenever we write an equation like (\ref{kk12}).
 This argument might be repeated later on several occasions in order to remind the reader that the presence of the extra terms, as alluded to,
 do not affect the analyticity properties of the amplitude.
   The polynomial boundedness
 of the scattering amplitude has been proved subsequently in the frameworks of general quantum field theories by Epstein, Glaser and Martin \cite{egm}.
 
 \bigskip
 
 \noindent
 {\bf 4.2.1 Scattering of} ${\bf n=0}$ {\bf Scalars States}
 
 \bigskip
 
 \noindent We  study the analyticity properties of the scattering amplitude of the KK zero modes of mass $m_0$.
 Thus two neutral massive scalars elastically scatter. The amplitude is 
 \bea
 \label{kk13}
 <k_d,k_c~out|k_b,k_a~in>-<k_d,k_c~in|k_b,k_a~in>=2\pi \delta^3({\bf k}_a+{\bf k}_b-{\bf k}_c-{\bf k}_d)F(k_a,k_b,k_c,k_d)
 \eea
 The notation is as follows: although we are considering scattering of identical, equal mass, neutral particles; it is convenient
 the label them. $k_a$ and $k_b$ are the four momenta of incoming particles ($a$ and $b$ respectively), $k_c$ and $k_d$ are the momenta of outgoing
 particles. All external particles are on mass shell. $F(k_a,k_b,k_c,k_d)$ is the scattering amplitude depending on Lorentz invariant
 variables  defined in (\ref{kk7}). We apply the LSZ reduction technique to derive the expression for the four point amplitude
 \bea
 \label{kk14}
 <k_d,k_c~out|k_b,k_a~in>=&& 4k^0_ak^0_b\delta^3({\bf k}_d-{\bf k}_b)\delta^3({\bf k}_a-{\bf k}_c) -{{i}\over{(2\pi)^3}}
 \int d^4x\int d^4x'e^{-i(k_a.x-k_c.x')}\nonumber\\&& K_xK_{x'}<k_d~out|R(x';x)|k_b~in>
 \eea
 where
 \be
 \label{kk15}
 R(x';x)=-i\theta (x_0-x_0')[\phi_a(x),\phi_c(x')]
 \ee
 We have reduced particles $a$ and $c$ in the above equation. $K_x$  and $K_{x'}$ are the Klein-Gordon (K-G) operators. This equation is similar to
 eq. (\ref{kk10}) except that we are now considering the zero modes of KK states. 
  We shall
 resort to the form of (\ref{kk14}) and abandon  that form of the amplitude where all four fields are reduced. The essential remarks are in order in the sequel:\\
 (i) The K-G operators in (\ref{kk14}) act on $R(x';x)$  in the following ways. When the first K-G operator acts on the $\theta$-function it will 
 give rise to a $\delta$-function and also it will produce a derivative of the $\delta$-function. K-G, with $x-$derivative acting on $\phi_a(x)$
 lead to the source current since the interacting field satisfies the equation of motion $(\Box_x+m_0^2)\phi_a(x)=j_a(x)$. As we have invoked the arguments of
 Symanzik earlier, $K_x$ and $K_{x'}$ acting on $R\phi_a(x)\phi_c(x')=Rj_a(x)j_c(x'')$ up to derivatives of $\delta$-functions. Their presence
 do not affect the analyticity properties of the amplitude.\\
 (ii) Thus the operation of the two K-G operators leaves us with R-product of two source currents and some extra terms whose nature have been noted
 earlier.  When we 
 write
 \be
 \label{kk16}
 (\Box_x+m_0^2)(\Box_{x'}+m_0^2)(R(x';x))=Rj_a(x)j_c(x')
 \ee
 it is understood that we are omitting presence of extra terms alluded to above. In a strict sense the equality is not valid.
  If we consider an arbitrary matrix element
 between $Rj_a(x)j_c(x')$ it can be brought to a form $Rj_a(x/2)j_c(-x/2)$ using the translation operations in those matrix elements as is well known.
  Thus we may introduce three (generalized) functions \cite{fr1,lehm1}
 \bea
 \label{kk17}
    F_R(q)=\int_{\infty}^{+\infty}d^4ze^{iq.z}\theta(z_0)<Q_f|[j_a(z/2),j_c(-z/2)]|Q_i>
    \eea
    \bea
    \label{kk18}
 F_A(q)=-\int_{\infty}^{+\infty}d^4ze^{iq.z}\theta(-z_0)<Q_f|[j_a(z/2),j_c(-z/2)]|Q_i>
 \eea
 and
 \be
 \label{kk19}
 F_C(q)= \int_{-\infty}^{+\infty}d^4ze^{iq.z}<Q_f|[j_a(x),j_c(x)]|Q_i>
 \ee
 The functions (\ref{kk17}) - (\ref{kk20}) are known respectively as the retarded, advanced and causal functions. Here  
 $|Q_i>$ and $|>Q_f$ are states which carry four momenta and these momenta are held fixed and we treat them as parameters as we shall 
 note in ensuing discussions. It is evident from above equations, (\ref{kk17}), (\ref {kk18}) and (\ref{kk19}), that
 \be
 \label{kk20}
 F_C(q)=F_R(q)-F_A(q)
 \ee
 Notice that $F_C$ is expressed as commutator of two currents. Then let us open up the commutator and thus we get $j_a(x)j_c(x')-j_c(x')j_a(x)$ between the 
 two states. 
 Let us introduce two complete set of physical
states: $\sum_n|{\cal P}_n{\tilde\alpha}_n><{\cal P}_n{\tilde\alpha}_n|={\bf 1}$ and
 $\sum_{n'}|{\bar{\cal P}}_{n'}{\tilde\beta}_{n'}><{\bar{\cal P}}_{n'}{\tilde\beta}_{n'}|={\bf 1}$.
Here $\{{\tilde\alpha}_n, {\tilde\beta}_{n'} \}$
stand for quantum numbers that are permitted for the physical intermediate states.
  However, the situation
 slightly different in this scenario in contrast to the well known prescriptions in the study of the analyticity domains of the amplitudes
 when we consider a theory with  only a single neutral scalar field. The intermediate
 states contribute from the entire Hilbert space which is direct sum of the disjoint Hilbert space each one of which is designated by a KK charge.
 Thus the intermediate physical states are such that their KK charge is zero in the scattering process of zero charge KK particles.  Here it is assumed that
 $|Q_i>$ and $|Q_f>$ are states with $n=0$.
 An intermediate state 
 could be a two particle or multiparticle state with total zero KK charge.
 Now we can express (\ref{kk19}) as
 \bea
\label{kk21}
&&\int d^4ze^{iq.z}\bigg[\sum_n\bigg(\int d^4{\cal P}_n<Q_f|j_a({z\over 2)}|{\cal P}_n{\tilde\alpha}_n
><{\cal P}_n{\tilde\alpha}_n|j_c(-{z\over 2})|Q_i>\bigg)\nonumber\\&& -
\sum_{n'}\bigg(\int d^4{\bar{\cal P}}_{n'}<Q_f|j_c(-{z\over 2})|{\bar{\cal P}}_{n'}{\tilde\beta}_{n'}>
<{\bar{\cal P}}_{n'}{\tilde\beta}_{n'}|j_a({z\over 2})|
Q_i>\bigg) \bigg]
\eea
If we use spacetime translation on each of the matrix elements in (\ref{kk21}) the $z$-dependence in the arguments of currents
 disappear i.e. they become $j_a(0)$ and $j_c(0)$; moreover  an energy momentum conserving $\delta$-function  appears. As a  consequence
 ${\cal P}_n={{(Q_i+Q_f)}\over 2}-q$ and  ${\bar{\cal  P}}_{n'}={{(Q_i+Q_f)}\over 2}+q$. Therefore,
\bea
\label{kk22}
&& F_C(q)=\sum_n\bigg(<Q_f|j_a(0)|{\cal P}_n={{(Q_i+Q_f)}\over 2}-q,{\tilde\alpha}_n><
{\tilde\alpha}_n,{\cal P}_n={{(Q_i+Q_f)}\over 2}-q|j_c(0)|Q_i>\bigg) \nonumber\\&&
-\sum_{n'}\bigg(<Q_f|j_c(0)|{\bar{\cal  P}}_{n'}={{(Q_i+Q_f)}\over 2}+q,{\tilde\beta}_{n'}><
{\tilde\beta}_{n'},{\bar{\cal P}}_{n'}={{(Q_i+Q_f)}\over 2}+q|j_a(0)|Q_i>\bigg)
\eea
 A few explanatory remarks are in order: (i) In the sum over intermediate states the lowest mass two particle state will be $4m_0^2$.
 All higher KK charged intermediate  states have higher thresholds since they should appear with zero-sum charges. Although we have 
 not specified $|Q_i>$ and $|Q_f>$  to have zero KK charge, for the problem at hand  (i.e. $\phi\phi\rightarrow \phi\phi$). These two states (when we consider scattering
 amplitude) will be particle states with zero KK charge. (ii) The second point to note is that entire tower of KK states will not
 contribute as intermediate states in the above expressions. In fact, for the scattering amplitude one term will be identified
 as absorptive part of the $s$-channel amplitude whereas the other term is the $u$-channel absorptive amplitude. Thus the
 contributions from KK towers will have finite number of terms. We shall provide a more convincing argument when we consider the
 elastic scattering of states with with nonzero KK charges (see section 5.1).\\
 Let us consider the contributions of the multiparticle states from zero-KK-charge sector (states belong to ${\cal H}_{q_n=0}$  space) 
 to the above expressions.
 We define
 \bea
\label{kk23}
2A_s(q)=&&\sum_{n'}\bigg(<Q_f|j(0)_a|{\bar{\cal P}}_{n'}={{(Q_i+Q_f)}\over 2}+q,{\tilde\beta}_{n'}>
\times \nonumber\\&&
<{\tilde\beta}_n',{\bar{\tilde P}}_n={{(Q_i+Q_f)}\over 2}+q|j_c(0)|Q_i>\bigg)=0
\eea
and 
\bea
\label{kk24}
2A_u=&&\sum_n\bigg(<Q_f|j_c(0)|{\cal P}_n={{(Q_i+Q_f)}\over 2}-q,{\tilde\alpha}_n>\times
\nonumber\\&&
<{\tilde\alpha}_n,{\cal P}_n={{(Q_i+Q_f)}\over 2}-q|j_l(0)|Q_i>\bigg)=0
\eea
Note that the Fourier transform of $F_C(q)$, ${\bar{F_C}}(z)$, vanishes outside the light cone as a consequence
of causality argument.  We note that with above definition of $A_u$ and $A_s$
\be
\label{kk24a}
F_C(q)= {{1}\over 2}(A_u(q)-A_s(q))
\ee
Moreover, $F_C(q)$ will vanish as function of $q$ wherever, both $A_s(q)$ and $A_u(q)$ vanish simultaneously.
We also remind that the  intermediate states are physical states and their four momenta lies in  the
forward light cone, $V^+$.
Consequently,  
\be
\label{kk25}
({{Q_i+Q_f}\over 2}+q)^2\ge 0,~~~({{Q_i+Q_f}\over 2})_0+q_0\ge 0
\ee
and
\be
\label{kk26}
({{Q_i+Q_f}\over 2}-q)^2\ge 0,~~~({{Q_i+Q_f}\over 2})_0-q_0\ge 0
\ee
These two equations, (\ref{kk25}) and (\ref{kk26}), imply that there ought to be minimum
mass parameters in each of the cases satisfying the conditions,
(i) $ ({{Q_i+Q_f}\over 2}+q)^2\ge {{\cal M}_+}^2$\\
 and
(ii) $({{Q_i+Q_f}\over 2}-q)^2\ge {{\cal M}_-}^2 $. If either of the  conditions (i) or (ii) is satisfied then one of the matrix element
will be nonvanishing and hence $F_C(q)\ne 0$; as 
$A_s(q)$ or $A_u(q)$ will not vanish.  
\\
The content of the above brief discussion is well known for quite sometime. The essential points to note is that, microcausality and Lorentz
invariance impose constraints on $F_C(q), F_R(q)$ and $F_A(q)$ in that the locations of the singularities in the complex
$q$-plane are identified. Therefore, 
 it is possible to derive the analyticity properties of the scattering  amplitude  starting from here in the following steps.
(i) The Jost-Lehmann-Dyson theorem allows us to find the location of singularities in the retarded function in the q-variable. Consequently,
the existence of small and large Lehmann ellipses is derived in the next step.\\
(ii) The fixed $t$ dispersion relations can be proved when  $t$  lies within Lehmann ellipse.\\
(iii) A host of known results can be derived for the field theory defined on $R^{3,1}\otimes S^1$.\\
(iv) For example, if we choose $Q_i=k_b$ and $Q_f=k_d$ then we shall derive expressions for the scattering
amplitude. Moreover, we shall be able to obtain expressions for $A_s$ and $A_u$ and relate them to absorptive
parts. It can be shown, in this case, that  there is a  region in the $q$ variable where $A_s(q)=A_u(q)$ for real $q$ and  $q$ lies in an unphysical
kinematical region. This is the coincidence region. The crossing symmetry of the amplitude is proved using the theory
of several complex variables.  Then the technique of enlarging the domain of holomorphy in the theory of several 
complex variables is  utilized  to prove the crossing symmetry. Our intent is not to address the issues related to crossing
symmetry in this article.\\  
The summary of this subsection, i.e. ${\bf{4.2.1}}$, is that for the zero mode field of a compactified field
theory ($R^{3,1}\otimes S^1$) the scattering amplitude satisfies the analytic properties of
known massive, neutral, scalar field theory as expected. This conclusion  was also reached by Khuri \cite{khuri2} in his model in the $n=0$ sector
of potential scattering.  

\bigskip

\noindent{\bf 4.2.2 Elastic Scattering of  Scalar $n=0$ State and $n\ne 0$ State}.

\bigskip

\noindent This subsection is devoted to study the elastic scattering of an $n=0$ particle with an 
$n\ne 0$ particle. We shall utilize the formalism developed in the Subsection 4.2 and take
into account the necessary modifications required for this case.\\
First thing to note is that  the initial states are the $n=0$ KK zero mode particle and the $n\ne 0$ KK particle which are
 denoted by $\phi_a$ and $\chi_b$ respectively with initial four momenta, $k_a$ and $p_b$. Similarly, the outgoing
 particles are denoted by $\phi_c$ and $\chi_d$ with final momenta, $k_c$ and $p_d$. The masses of $\phi$ and $\chi$
 respectively are $m_0^2$ and $m_n^2=m_0^2+ {{n^2}\over{R^2}}$ and we are focusing only on the elastic process.  The
 source currents associated with the interacting fields $\phi(x)$ and $\chi(x)$ are denoted by $j(x)$ and $J(x)$ respectively. 
 Note that the initial and the final $c.m.$ momenta, ${\bf k}$, are the same for this reaction.\\
 The amplitude is defined as 
\bea
 \label{kk27}
 <p_d,k_c~out|p_b,k_a~in>-<p_d,k_c~in|p_b,k_a~in>=2\pi \delta^3({\bf k}_a+{\bf p}_b-{\bf k}_c-{\bf p}_d)F(k_a,p_b,k_c,p_d)
 \eea
\bea
 \label{kk28}
 <p_d,k_c~out|p_b,k_a~in>=&& 4k^0_ak^0_b\delta^3({\bf p}_d-{\bf p}_b)\delta^3({\bf k}_a-{\bf k}_c) -{{i}\over{(2\pi)^3}}
 \int d^4x\int d^4x'e^{-i(k_a.x-k_c.x')}\nonumber\\&& K_xK_{x'}<k_d~out|R(x';x)|k_b~in>
 \eea
Here we have chosen to reduce the states $\phi_a$ and $\phi_c$ and the two states with $n\ne 0$ are not reduced. Thus $R(x';x)$ has the same
definition as in the previous subsection, 4.2.1, (see (\ref{kk15})).  Therefore, we may write the matrix element as before and introduce
the three analogous functions, $F_{R}(q),~F_A(q)$ and $F_C(q)$ where the the commutator of the source current $[j_a(z/2),j_c(-z/2)]$  is
sandwiched between two fixed arbitrary states $|Q_i>$ and $|Q_f>$. When we discuss the properties of the scattering
amplitude we should keep in mind that the unreduced initial and final states states have $n\ne 0$. \\
{\it Remark}: It is convenient to assign states
$|Q_i>$ and $|Q_f>$ nonzero $n$ quantum number but same value of $n$ since the total  KK charge is conserved for the 
initial and final configurations;  and note that $\phi$ states carry $n=0$ KK charge.
In view of above remarks let us examine the structure of $F_C(q)$ matrix element. As before, we open up the commutator
$[j_a(x),j_c(x')]$ introduce a complete set of states between the products of the two current.  Let us write down the relevant equations
for the problem at hand  
 \bea
\label{kk29}
&&\int d^4ze^{iq.z}\bigg[\sum_n\bigg(\int d^4{\cal P}_n<Q_f|j_a({z\over 2)}|{\cal P}_n{\tilde\alpha}_n
><{\cal P}_n{\tilde\alpha}_n|j_c(-{z\over 2})|Q_i>\bigg)\nonumber\\&& -
\sum_{n'}\bigg(\int d^4{\bar{\cal P}}_{n'}<Q_f|j_c(-{z\over 2})|{\bar{\cal P}}_{n'}{\tilde\beta}_{n'}>
<{\bar{\cal P}}_{n'}{\tilde\beta}_{n'}|j_a({z\over 2})|
Q_i>\bigg) \bigg]
\eea
If we use spacetime translation on each of the matrix elements in (\ref{kk21}) the $z$-dependence in the arguments of currents
 disappear i.e. they become $j_a(0)$ and $j_c(0)$; moreover  an energy momentum conserving $\delta$-function  appears
 after the $d^4z$ integration. As a  consequence, 
 ${\cal P}_n={{(Q_i+Q_f)}\over 2}-q$ and  ${\bar{\cal  P}}_{n'}={{(Q_i+Q_f)}\over 2}+q$. Therefore,
\bea
\label{kk30}
&& F_C(q)=\sum_n\bigg(<Q_f|j_a(0)|{\cal P}_n={{(Q_i+Q_f)}\over 2}-q,{\tilde\alpha}_n><
{\tilde\alpha}_n,{\cal P}_n={{(Q_i+Q_f)}\over 2}-q|j_c(0)|Q_i>\bigg) \nonumber\\&&
-\sum_{n'}\bigg(<Q_f|j_c(0)|{\bar{\cal  P}}_{n'}={{(Q_i+Q_f)}\over 2}+q,{\tilde\beta}_{n'}><
{\tilde\beta}_{n'},{\bar{\cal P}}_{n'}={{(Q_i+Q_f)}\over 2}+q|j_a(0)|Q_i>\bigg)
\eea
 Remark: (i) We have introduced a complete set of physical states as we had in (\ref{kk21}). We recall that we have assigned nonzero ( and the same )
 KK charges to the two states in the matrix element. Moreover,   KK charge is an additive conserved quantum number. (ii) Therefore, the admissible
 physical intermediate states appearing in equations (\ref{kk29}) and (\ref{kk30}) must carry $n$-units of KK charge. (iii)
 There are several possibilities: (a) A zero-mode state  ($n=0$) with a state of single KK state carrying $n$-unit of KK charge.
 The attributes of these intermediate states can be understood from (\ref{kk9}). (b) Several
 combinations of KK tower states where some may carry  $n=0$ KK charge; however, sum total of the charges of the multiparticle states
 must add up to 'n'. We would like to draw attention to the fact that such contributions arise due to the
 presence of KK tower of states. However, there will be only finite number of such intermediate states in the sum since these multi particle
 states are physical and they conserve energy momentum. This argument is intuitively sound. We mention in passing that our main purpose
 is to investigate the analyticity property of the amplitude for elastic scattering of states carrying nonzero KK charges. The problem at hand 
  (in this subsection) is that we have one particle with KK charge $n=0$ and other one has $n\ne 0$.\\
  We may correspondingly define $A_u(q)$ and $A_s(q)$ in analogy with equations (\ref{kk23}) and (\ref{kk24}).
  We intend to argue, in this case, $F_C(q)$ vanishes when each of the two terms in (\ref{kk30}) vanish for certain values of $q$. 
  In order that  $F_C(q)$  is nonzero, one of the matrix
  elements $A_u(q)$ or  $A_s(q)$  should be nonzero. On this occasion, we also have constraints
  \be
\label{kk31}
({{Q_i+Q_f}\over 2}+q)^2\ge 0,~~~({{Q_i+Q_f}\over 2})_0+q_0\ge 0
\ee
and
\be
\label{kk32}
({{Q_i+Q_f}\over 2}-q)^2\ge 0,~~~({{Q_i+Q_f}\over 2})_0-q_0\ge 0
\ee
as was derived in the previous case. We may follows the  arguments presented in $4.2.1$  that the analyticity properties
of this amplitude can be studied even when some of the particles carry  a nonzero KK charge. There is not much of a 
complication in the elastic scattering of two unequal mass particles as is the case here. Therefore, the conclusions
drawn at the end of subsection $4.2.1$ hold good. Moreover, we are allowed to write a fixed $t$
dispersion relation and do not foresee any difficulty.  However, to do so, we have to prove the existence of the 
Lehmann ellipses which does not seem to be a very challenging problem.Thus the forward dispersion relation can be proved in the  case analysed
in this subsection
\be   
 \label{kk33}
 (n=0)+(n\ne 0) \rightarrow (n=0)+(n\ne 0)
 \ee
 {\it This scattering process was not addressed by Khuri}. The conclusion of this subsection is that the scattering amplitude, considered
 in this subsection, will satisfy forward dispersion relation. In fact the analyticity hold for nonforward direction
 as long as $t$ lies inside the corresponding Lehmann-Martin ellipse.
  
  \bigskip
  
  \noindent {\bf 5.  Elastic Scatting of States with nonzero Kaluza-Klein Charges}
  
  \bigskip
  
  \noindent  The elastic scattering of two particles carrying nonzero Kaluza-Klein charges are studied here. We repeat
  some of the assumptions alluded to in the beginning. These are neutral, massive, scalar particles.
  They are termed as neutral in the sense that they do not carry electric charge (more generally they do not carry any charge
  associated due to a local gauge symmetry). Of course, the states considered in this section, carry the KK charges.
   These particles are
  stable and there are no bound states in the theory. If we consider two incoming particles with KK charges $m$ and $n$
  their masses are respectively $m_0^2+{{m^2}\over{R^2}}$ and  $m_0^2+{{n^2}\over{R^2}}$. We assume, without any loss of generality,
  that each of the particles carries same KK charge, $n>0$, in order to simplify the computations and  consequently, all four participating particles are of equal mass. 
  It will not affect any of the conclusions if we considered elastic scattering of two KK particles with different charges 
  as will be evident later. In fact, the prescription laid down so far is adequate to handle
  elastic scattering of particles with unequal KK charges and hence unequal masses. Indeed, Khuri \cite{khuri2} has derived the 
  result for the case $(n)+(n)\rightarrow (n)+(n)$ for forward scattering.  He had chosen the {\it special case}  of $n=1$ in order to demonstrate 
  through a counter example that the dispersion relation is violated for that case. Our goal is to study the analyticity
  property of forward amplitude in this context. The initial incoming pair of particles are denoted by $\chi_a$ and $\chi_b$  and they carry four momenta 
  $p_a$ and $p_b$ respectively. The outgoing particle are $\chi_c$ and $\chi_d$ and carry  four momenta $p_c$ and $p_d$.
  Our first step is to define the scattering amplitude for this reaction and proceed systematically 
  \bea
 \label{nn1}
 <p_d,p_c~out|p_b,p_a~in>=&& 4p^0_ap^0_b\delta^3({\bf p}_d-{\bf p}_b)\delta^3({\bf p}_a-{\bf p}_c) -{{i}\over{(2\pi)^3}}
 \int d^4x\int d^4x'e^{-i(p_a.x-p_c.x')}\nonumber\\&& {\tilde K}_x{\tilde K}_{x'}<p_d~out| R(x';x)|p_b~in>
 \eea
  where
\be
 \label{nn2}
  R(x';x)=-\theta (x_0-x_0')[\chi_a(x),\chi_c(x')]
 \ee  
  and ${\tilde K}_x=(\Box+m_n^2)$. We let the two K-G operators act on ${\bar R}(x;x')$  in the VEV and resulting
  equation is
  \bea
  \label{nn3}
 <p_d,p_c~out|p_b,p_a~in>=&&<p_d,p_c~in|p_b,p_a~in> 
  -{{1}\over{(2\pi)^3}}
 \int d^4x\int d^4x'e^{-i(p_a.x-p_c.x')}\nonumber\\&&
  <p_d|\theta(x_0'-x_0)[J_c(x'),J_a(x)]|p_b>
  \eea
  Here $J_a(x)$ and $J_c(x')$ are the source currents associated with the fields $\chi_a(x)$ and $\chi_c(x')$ respectively.
  We arrive at (\ref{nn3}) from (\ref{nn1}) with the understanding that the $R.H.S.$ of (\ref{nn3}) contains additional
  terms as discussed earlier; however, these terms do not affect the study of the analyticity properties of the amplitude.
  We have mentioned in the previous section that we shall explore the consequences unitarity in this section and the 
  purpose will be transparent presently.
  Let us define the $\bf T$-matrix as follows:
  \be
  \label{nn4}
  {\bf S}={\bf 1}-i{\bf T}
  \ee 
  The unitarity of the S-matrix, ${\bf {SS^{\dagger}}}={\bf {S^{\dagger}S}}={\bf 1}$ yields
  \be
  \label{nn5}
  ({\bf{T^{\dagger}}}-{\bf{T}})=i{\bf{T^{\dagger}T}}
  \ee
  In the present context,  we consider the matrix element  for the reaction $a+b\rightarrow c+d$. Note that on
  $L.H.S$ of ({\ref{nn5}) it is taken between ${\bf{T^{\dagger}-T}}$. We introduce a complete set of physical states
  between ${\bf {T^{\dagger}T}}$.   For the elastic case with all particles of KK charge, $n$, the unitarity relation is
  \bea
  \label{nn6}
  <p_d,p_c~in|{\bf{T^{\dagger}}}-{\bf T}|p_b,p_a~in>=i\sum_{{\cal N}}<p_d,p_c~in|{\bf{T^{\dagger}}}|{\cal N}><{\cal N}|{\bf T}|p_b,p_c~in>
  \eea
  The complete of states  stands for $|{\cal N}> =| p_N;{\tilde\alpha}_N>$; $p_N$ is the momentum of a physical state and 
  ${\tilde\alpha}_N$ stands for all other quantum numbers and some times
  we do not explicitly display its presence in the  matrix elements.  The unitarity relation reads,
  \bea
  \label{nn7}
  T^*(p_a,p_b;p_c,p_d)-T(p_d,p_c;p_b,p_a)=2\pi i\sum_ N\delta(p_d+p_c-p_N)T^*(N;p_c,p_d)T(N;p_b,p_a)
  \eea
  We arrive at an expression like the second term of the $R.H.S$ of (\ref{nn1}) after reducing two fields. If we reduce a single field
  as the first step (as is worked out in text books) there will be a single K-G operator acting on the field and eventually
  we obtain matrix element of only a single current.  The $R.H.S.$ of (\ref{nn7}) has matrix element like (for example)
  $p_a+p_b\rightarrow p_N$. Thus we can express it as \footnote{ We  adopt the arguments and procedures of Gasiorowicz in these derivations}
    \cite{gasio}
  \be
  \label{nn8}
  \delta(p_N-p_a-p_b)T(N:p_b,p_a)= (2\pi)^{3/2}<N~out|J_a(0)|p_b>\delta(p_N-p_a-p_b)
  \ee
  After carrying out the computations we arrive at
  \bea
  \label{nn9}
  T(p_d,p_c;p_b,p_a) -T^*(p_d,p_c;p_b,p_a)=&&\sum_ N\bigg[\delta(p_d+p_c-p_N)T(p_d,p_c;N)T^*(N;p_b,p_a)-\nonumber\\&&
  \delta(p_a-p_c-p_N)T(p_d,-p_c;N)T^*(p_d,-p_c;N) \bigg]
  \eea
  Let consider the scattering amplitude for the reaction under considerations.
  \bea
  \label{nn10}
  F(s,t)=i\int d^4xe^{i(p_a+p_c).{{x}\over 2}}\theta(x_0)<p_d|[J_a(x/2),J_c(-x'/2)]|p_b>
  \eea
  We define below the imaginary part of this amplitude, $F(s,t)$ and evaluate it
  \bea
  \label{nn11}
  Im~F(s,t)=&&{{1}\over{2i}}(F-F^*)\nonumber\\&&
  ={{1}\over2}\int d^4xe^{i(p_a+p_c).{{x\over 2}}}<p_d|[J_a(x/2),J_c(-x/2)]|p_b>
  \eea
  Note that $F^*$ is invariant under interchange $p_b\rightarrow p_d$ and also $p_d\rightarrow p_b$; moreover, $\theta(x_0)+\theta(-x_0)=1$.
  We open up the commutator of the two currents in (\ref{nn11}). Then introduce a complete set of physical states $\sum_{\cal N}|{\cal N}><{\cal N}|=1$. Next we
  implement  translation operations in each of the (expanded) matrix elements to  express  arguments of each current as  $J_a(0)$ and $J_c(0)$
  and finally integrate over $d^4x$  to get the $\delta$-functions. As a consequence  (\ref{nn11}) assumes the form
  \bea
  \label{nn12}
  F(p_d,p_c;p_b,p_a)-F^*(p_b,p_a;p_c,p_d)=&&2\pi i\sum_N\bigg[\delta(p_d+p_c-p_N)F(p_d,p_c;N)F^*(p_a,p_b;N)\nonumber\\&&
  -\delta(p_a-p_c-p_N)F(p_d,-p_a;N)F^*(p_b,-p_c;N)\bigg]
  \eea
  This is the generalized unitarity relation  and  all the external particles are on the mass shell. Notice that the first term on the $R.H.S$ of the
  above equation is identical in form to the $R.H.S.$ of (\ref{nn9}); the unitarity relation for ${\bf T}$-matrix. The first term in (\ref{nn12})  has
  the following interpretation: the presence of the $\delta$-function and total energy momentum conservation implies
  $p_d+p_c=p_N=p_a+p_b$. We identify it as the $s$-channel process $p_a+p_b\rightarrow p_c+p_d$.\\
  Let us examine the second term of (\ref{nn12}).  Recall that the unitarity holds for the $S$-matrix when all external particles are on shell
  (as is true for the $T$-matrix). The presence of the $\delta$-function in the expression ensures that the intermediate physical states will contribute for
  \be
  \label{nn13}
  p_b+(-p_c)=p_N=p_d+(-p_a)
  \ee
  The masses of the intermediate states must satisfy
  \be
  \label{nn14}
  {\cal M}_N^2=p_N^2=(p_b-p_c)^2
  \ee  
  It becomes physically transparent if we choose the Lorentz frame where  particle  $'b'$ is at rest i.e. $p_b=(m_b, {\bf 0})$; thus
  \be
  \label{nn15}
  {\cal M}_N^2=2m_b(m_b-p_c^0),~~p_c^0>0
  \ee
  since $m_b=m_c$ and $p_c^0={{\sqrt{m_c^2+{\bf p}_c^2}}}={{\sqrt{m_b^2+{\bf p}_c^2}}} $;  ${\cal M}_N^2<0$ in this case.
  We recall that all particles carry the same KK charge $n$ and hence the mass is $m_b^2=m_n^2=m_0^2+{{n^2}\over{R^2}}$. The intermediate state
  must carry that quantum number. In conclusion, the second term of (\ref{nn12}) does not contribute to the $s$-channel reaction. There is
  an important implication of the generalized unitarity equation: Let us look at the crossed channel reaction
  \be
  \label{nn16}
  p_b+(-p_c)\rightarrow p_d+(-p_a);~~ -p_a^0>0,~ and~ -p_c^0>0
  \ee
  Here $p_b$ and $p_c$ are incoming (hence the negative sign for $p_c$) and $p_d$ and $p_a$ are outgoing. The second matrix element
  in (\ref{nn12}) contributes to the above process in the configurations of the four momenta of these particles as noted in (\ref{nn16});
   whereas the first term in that equation does not if we follow the arguments for
  the $s$-channel process. \\
  {\it Remark}: We notice the glimpses of crossing symmetry here. As we have argued earlier in subsection 4.2.1. Indeed, the starting
  point will be to define $F_C(q)$ and look for the coincidence region. Notice that  $q$ is related to physical momenta of external particles when
  $|Q_i>$ and $|Q_f>$ are identified with  the momenta of the 'unreduced' fields. Indeed, we could proceed to prove crossing  symmetry for the 
  scattering process; however,  it is not our present goal.\\
   {\it Important observations are in order}: \\
    (i) We could ask whether entire Kaluza-Klein tower  of states
  would appear as intermediate states in the unitarity equation \footnote {I would like to thank Luis Alvarez Gaume for raising this issue.}.
  It is obvious from the unitarity equation (\ref{nn12}) that for the $s$-channel process, due to the presence of the energy momentum
  conserving $\delta$-function, $p_n^2={\cal M}_n^2= (p_a+p_b)^2$; consequently, not all states of the infinite KK tower will contribute
  to the reaction in this,   ($s$), channel. Therefore the sum would terminate after finite number of terms, even for very large $s$ as long as it is finite.
   Same argument also holds for the crossed channel process.  
  \\
  (ii) We could proceed to prove crossing symmetry from this point. However, our goal, in this investigation, is to specifically examine whether
  the forward scattering amplitude satisfies in the present case. Thus we focus on the forward amplitude. It is convenient to adopt the prescriptions 
  of Symanzik \cite{kurt} to this end.
  
  \bigskip
    
  \noindent {\bf 5.1 Dispersion Relation for Forward Scattering Amplitude}

\bigskip

\noindent  We intend to study the analyticity property of the elastic scattering amplitude in the forward direction in this subsection.
This is of paramount interest to us  since it was shown by Khuri, in the case of potential scattering in a quantum mechanical model with $R^3\otimes S^1$
geometry, that the elastic forward scattering amplitude $T(k,k; n,n)$ does not have analyticity property in contrast to the  case with potentials
with noncompact coordinates. We utilize the formalism developed in Section 4 and in this section. The configuration for the forward scattering amplitude is
$p_c=p_a$ and $p_d=p_b$. Moreover, all the particles are scalars and of equal mass, $m^2_n=m_0^2+{{n^2}\over{R^2}}$. Recall that the amplitude
is a function of the Lorentz invariant variables, $s$ and $t=0$. The forward amplitude is
\bea
\label{nn17}
F(p_b,p_a;p_b,p_a)=\int d^4xe^{ip_a.x}(\Box_x+m_n^2)^2<p_b|R\chi_a(x)\chi_a(0)|p_b>
\eea
leading to 
\bea
\label{nn17a}
F(p_b,p_a;p_b,p_a)=\int d^4xe^{ip_a.x}<p_b|RJ_a(x)J_a(0)|p_b>
\eea

We remind that in the nonforward case the reader that we had an exponential term $exp (i(p_a+p_c).x/2)$ and there were two K-G operators
$K_xK_{x'}$ acting on two currents $RJ_c(x)J_a(0)$ while we considered nonforward amplitude.\\
We shall adopt the procedure of Symanzik who proved the dispersion relations for forward scattering amplitude of pion-nucleon scattering. Indeed,
this is a simpler case of equal mass scattering. Alternatively, one could adopt mathematically more rigorous formulation of Bogoliubov; however,
we  have resorted to the approach of \cite{kurt}. We go to a frame where particle $'b'$ is at rest, ${\bf p}_b=0$. Introduce the Lorentz invariant
variable, $\omega= {{p_a.p_b}\over{m_n}}$ which is the incident energy of $'a'$ in this frame. The amplitude, (\ref{nn17}) takes the form
\bea
\label{nn18}
F(p_b,p_a:p_b,p_a)=i\int_0^{\infty} dt \int d^3{\bf x}e^{ip_a^0x^0-i{{\sqrt{(p_a^0)^2-m_n^2}}}{\hat{\bf e}}.{\bf x}} {\tilde f}({\bf x}, x_0)
\eea 
where ${\hat{\bf e}}$ is the unit vector along direction of the three momentum ${\bf p}_a$. We can read off expression for
${\tilde f}$ from (\ref{nn17}); at this stage it is sufficient  to note that that ${\tilde f}({\bf x}, x_0)$ vanishes unless $x_0>|{\bf x}|$ due to microcausality 
arguments (see more discussions below). We may carry out the angular integration and the resulting expression is
\be
\label{nn19}  
F(p_b,p_a;p_b,p_a)=\int_0^{\infty} {\cal F}(\omega ,r)dr
\ee
where
\bea
\label{nn20}
{\cal F}(\omega, r)=&&{4\pi ir^2}{{sin{\sqrt{\omega^2-m_n^2}r}e^{i\omega r}}\over{{\sqrt{\omega^2-m_n^2}r}}} \times \nonumber\\&&
\int_r^{\infty} dte^{i\omega(r-t)}<p_b|[J_a(x),J_a(0)]|p_b>
\eea
Notice that ${\cal F}(\omega,r)$ is analytic function of $\omega$ in the upper half $\omega$-plane, i.e. for complex $\omega$,
$Im~\omega \ge 0$.  The following features of the $R.H.S.$ of (\ref{nn20}) are noteworthy.\\
 (i)  It appears that there might be a branch point at $\omega=\pm m_n$. However, note that there is really no branch point
at $\omega=\pm m_n$ since, as we observe,   ${sin{\sqrt{\omega^2-m_n^2}}re^{i\omega r}\over{r{\sqrt{\omega^2-m_n^2}}}}$ is an even function
of $r{\sqrt{\omega^2-m_n^2}}$. (ii) When $\omega<m_n$ we might apprehend the about the large $r$ behavior of  $sin{\sqrt{\omega^2-m_n^2}}r$
in the complex upper half $\omega$ plane; however, the presence
of $e^{i\omega r}$ dispels any such doubt. We have used, loosely, the equality $(\Box_x+m_n^2)(R\chi_a(x)\chi_a(0))=RJ_a(x)J_a(0)$. However,
as alluded to earlier the equality is up to the presence of finite number of derivatives of $\delta$-functions. Thus the amplitude, in the case of forward scattering,
 might have additional terms on $R.H.S.$ which are polynomials in $s$, in this Lorentz frame $\omega$
 (see more discussions on this point
later).\\
In what follows, we shall incorporate the essential arguments of Symanzik and give some modified steps of his derivation rather
than repeat the entire technical details he provided for the $\pi-N$ scattering. 
Let us assume that large $\omega$ behavior of the forward amplitude requires no subtractions and therefore, for large $\omega$,  the integral over
$\omega$ vanishes in the dispersion integral expressed as
\bea
\label{nn21}
{\cal F}(\omega, r)={{1\over\pi}}\int_{\infty}^{+\infty}{{{\rm Im}~{\cal F}(\omega',r)}\over{\omega'-\omega-i\epsilon}}d\omega'
\eea
We conclude from the expression for ${\cal F}(\omega,r)$, (\ref{nn20}), that ${\rm Im}~{\cal F}(r,\omega)=-{\rm Im}~{\cal F}(-\omega,r)$; 
consequently, the amplitude is an even function of energy, $\omega$. We may rewrite (\ref{nn21}) as
\bea
\label{nn22}
{\cal F}(\omega, r)= {{1}\over \pi}\int_0^{+\infty}  {\rm Im}~{\cal F}(\omega',r) \bigg[{{1}\over{\omega'-\omega-i\epsilon}}+
{{1}\over{\omega'-\omega+i\epsilon}} \bigg]
\eea
Note, from the definition of ${\cal F}(\omega,r)$, (\ref{nn21}), that the integration over $r$ is to be completed if we
want to derive dispersion relation for the amplitude $F(p_b,p_a;p_b,p_a)$.  Thus there arises the issue of interchange of integrations
over $r$ and $\omega$. Eventually, we are to compute the forward absorptive amplitude, (${\rm Im}~F$), as given in (\ref{nn20}). We
arrive at the following expression from (\ref{nn17a})
\be
\label{nn23}
{\rm Im}~ F(p_b,p_a;p_b,p_a)={{1\over 2}}\int d^4xe^{p_a.x}<p_b|[J_a(x),J_b(0)]|>
\ee
The angular integration for $e^{ip_a.x}$ can be carried out in the Lorentz frame of our choice leading to an expression analogous
to (\ref{nn20}). Thus
\bea
\label{nn24}
{\rm Im}~ F(p_b,p_a;p_b,p_a)=&&{{1}\over 2}\int dr4\pi r{{{sin\sqrt{\omega^2-m_n^2}r}}\over{\sqrt{\omega^2-m_n^2}}} \times \nonumber\\&&
\int_{-\infty}^{+\infty}e^{i\omega t}<p_b|[J_a(x),J_a(0)]|p_b>dt
\eea
We recall that in deriving the generalized unitarity relation for the nonforward amplitude, we had opened up the commutator
of the two currents and we inserted complete set of states in the products of currents in the each term. Notice that in the present
context, the initial and the final states are identical; this is the route to derive the optical theorem. However, our goal is different here. 
As has been our practice earlier, we use translation operation to get rid of the $x$-dependence in the argument of one of the current.
Consequently, a factor of $e^{i{\bf p}_N.{\bf x}}$ or $e^{-i{\bf p}_N.{\bf x}}$ would appear where ${\bf p}_N$ is the momentum of the
physical intermediate state ( we are still in the same Lorentz frame ).  Therefore, as before, the angular integration can be carried out
resulting the factors depending on $r$ only
\bea
\label{nn25}
{\rm Im}~{\cal F}(\omega, r)=&&2\pi r{{sin{\sqrt{\omega^2-m_n^2}}r}\over{{\sqrt{\omega^2-m_n^2}}}}\sum_N{{sin|{\bf p}_N|r}\over{|{\bf p}_N|r}}
\nonumber\\&&|<p_b|J_a(0)|N>|^2\bigg[\delta(\omega+p_N^0-m_n)-\delta(\omega+m_n-p_N^0)\bigg]
\eea
Here $p_N$ is the four momentum of the intermediate state (recall we used completeness relation $ \sum_{\cal N}|{\cal N}><{\cal N}|={\bf 1}$
where ${\cal N}$ stood for intermediates permitted by energy momentum conservation and collection of all discrete quantum numbers
such that KK charge is conserved). The same logic applies here. Notice that the source current $J_a$ carries KK charge of  $n$ units.
Thus the  state $J_a(0)|n>$ has to be such that its KK charge is also $n$ unit since $<p_b|$ carries $n$ unit of KK charge.
If  we carry out $d^4x$ integration in (\ref{nn23}) then we shall get an expression
similar to (\ref{nn11}), now in the forward direction.
\bea
\label{nn26}
{\rm Im}~ F(p_b,p_a;p_b,p_a)={1\over 2}(2\pi)^4\sum_N|<p_b|J_a(0)|p_b>|^2\bigg[\delta^4(p_b+p_a-p_N)-\delta^4(p_b-p_a+p_N)\bigg]
\eea
It is worthwhile to draw the similarities of the present investigation with the work of Symanzik \cite{kurt} who proved dispersion relation for
the forward scattering amplitude of pion-nucleon scattering. We shall discuss the situation where there is departure of the present work
from that of \cite{kurt}. The isospin quantum numbers of pion and nucleon were not accounted for and both the 
particles were taken to be spinless.  Consequently, the complications due to the nucleon spin were not encountered in that  situation.  Thus the problem
was reduced to the study of the  scattering of two unequal mass spinless particles. However, the nucleon was assigned an additive conserved quantum number.
Symanzik reduced the two nucleons of initial and final states when he implemented LSZ formalism. In the process of computation of the
amplitude, while introducing complete set of states between product of currents in the matrix elements all the conservation laws were
accounted for. The dispersion integral was obtained keeping in mind the issues alluded to above.  At that juncture, he argued that, at high energy,
the amplitude will have, at most, polynomial growth. The proof of Jin-Martin \cite{jml} theorem that the elastic amplitude needs no more than
two subtractions was not known in 1957. Moreover,  the general results of Epstein, Glaser and Martin \cite{egm} appeared much later.\\
In order to contrast our work with Symanzik's, we note the important feature that all the particles carry additive  discrete quantum charge, $n$, and these
are equal mass bosons. Consequently, once we reduce $'a'$ and $'c'$ ,see (\ref{nn26}), (before considering the case of forward scattering) the two
unreduced states ($'b'$ and $'d'$)  each carry $n$-units of charge.  Thus the intermediate states sandwiching between one current with either
$|b>$ or $<d|$ must respect  the  desired conservation laws. We have not proved the Jin-Martin  \cite{jml} bound in this article. Nonetheless it will suffice
if the amplitude has at most a polynomial growth in $s$ for large s.  We can write a subtracted dispersion relation in such a case.  Notice that
in the case of pion-nucleon scattering, there is a stable nucleon carrying the baryon number and the one-particle intermediate state is the nucleon.
For the case at hand the intermediate state is to carry two units of $n$-charge. If the particles of mass ${\sqrt{m_0^2+{4n^2/{R^2}}}}$ appears as a pole in the amplitude,
we can account for this pole as the presence of nucleon pole was taken care of in the pion-nucleon process.\\
Notice that, keeping the above remarks in mind, the rest of the computation  could be developed in parallel to the case of the pion-nucleon scattering.
The interchange of $r$ and $\omega$ integrations in defining the ${\rm Im}~{\cal F}$ in the dispersion integral can be justified
if we adopt arguments of Symanzik and Gasiorowicz \cite{kurt,gasio}. On the other hand, Bogoliubov introduced a prescription to obtain the dispersion relation for the scattering
amplitude.  It is now obvious  that if we follow either of the procedures Symanzik \cite{kurt}  or Bogoliubov \cite{bogo} the forward scattering amplitude for
elastic the process 
\be
\label{nn27}
p_a(n)+p_b(n)\rightarrow p_a(n)+p_b(n)
\ee
  satisfies dispersion
relation in $\omega$ and hence in $s$.  In view of the above discussions, the forward scattering amplitude might admit a pole with KK charge of $2n$ units. Thus
the presence of possible pole term does not affect the ensuing argument about writing a dispersion relation. Now,  if we resort to 
Mandelstam variables $s$ and $u$ and recall $u=4m_n^2-s$ for $t=0$, the dispersion relation can be written down in the familiar
form
\bea
\label{nn27}
{\rm F(s,t=0)}={{1\over{\pi}}}\int_{s_{thr}}^{\infty}ds'{{{\rm Im}~F(s',t=0)}\over{s'-s}}+{{1\over{\pi}}}\int_{u_{thr}}^{\infty}du'{{{\rm Im}~F(u',t=0)}\over{u'-u}}
 \eea
Therefore, 
it  is demonstrated under the stated assumptions that the imaginary part of the amplitude is tends to zero as $s\rightarrow \infty$
(i.e. sufficiently convergent) the forward amplitude satisfies an unsubtracted 
dispersion relation.
If the amplitude  i is polynomially bounded as $s$ tends to large values, we can introduce
finite number of subtractions as the analyticity property will continue to hold. In case we invoke Jin-Martin \cite{jml} upper bound
for the amplitude then the forward amplitude will need {\it no more} than {\it two} subtractions (see discussion for more details).\\
{ The important conclusion  of our investigation is that the forward scattering amplitude for elastic scattering of KK states with
$n> 0$, in  a  scalar field theory defined in a compactified spacetime, i.e. $R^{3,1}\otimes S^1$; satisfies dispersion relation in $s$   
in the forward direction. This conclusion  is similar  to the  case of a massive, neutral, scalar field theory defines in a flat four dimensional spacetime, $R^{3,1}$.
\\
Our
main conclusion may be stated as a theorem.\\
{\it Theorem:  A massive neutral scalar field theory is defined on $R^{4,1}$  and the manifold is compactified to $R^{3,1}\otimes S^1$ subsequently.
  The spectrum of the
states are obtained by the Kaluza-Klein dimensional reduction. The forward elastic scattering amplitude for scattering of the Kaluza-Klein states
 on the manifold $R^{3,1)}\otimes S^1$ satisfies dispersion relations.}  
\bigskip

\noindent{\bf 6 Summary and Discussions}

\bigskip

\noindent We summarize our results in this section and discuss their implications.  Our principal goal of this article  is  to study
the analyticity property of the forward scattering amplitude for a five dimensional scalar field theory which is compactified
to $R^{3,1}\otimes S^1$. The interest in this problem arose from a work of Khuri \cite{khuri2} in potential scattering in a spatial geometry of 
$R^3\otimes S^1$. He adopted the Green's function technique and employed perturbation theory in order to compute the scattering amplitude. 
The main conclusion was
that for a class of Yukawa-type potentials, the forward scattering amplitude failed to satisfy the dispersion relation in the
second order for the case when the discrete quantum number, $n$, associated with the periodic coordinate of $S^1$, is
nonzero. Indeed, there were some concerns if dispersion relation is invalidated in relativistic quantum field theories.\\
I have worked in the axiomatic frameworks of LSZ (Lehmann-Symanzik- Zimmermann).
I considered a neutral,  scalar field of mass $m_0$  in $D=5$ and compactified the spacetime to $R^{3,1}\otimes S^1$. The resulting
theory is Lorentz invariant in four dimensional spacetime. It has a massive, neutral scalar field of mass $m_0$ ( {\it mass of the zero mode}) in
addition to a tower of KK states. I presented a brief outline of Khuri's result in Section 2. I developed the systematic prescription
to study the field theory in the $D=4$ spacetime with a compact dimension. The sections 4, 4.2.1, 4.2.2 were devoted to
setting up the frame work. The first case was to consider elastic scattering of $n=0$ particles. This problem has been studied
long ago in the frameworks of axiomatic field theory. The only departure, from the standard case, is the presence of KK towers. We argued qualitatively that the entire
tower of KK states do not appear as intermediate states when we derive spectral representations for absorptive parts. It is to be
noted that, for this case, we can derive the analyticity properties rigorously for the forward amplitude as well as for the nonforward amplitude.
 Moreover, we can prove
the existence Lehmann ellipses and thus write down fixed $t$ dispersion relation as long as $|t|$ lies inside the Lehmann-Martin ellipse.
In fact, Khuri \cite{khuri2}, in his analysis,  concluded that the amplitude for $n=0$ sector satisfies analyticity properties in his potential model.
Next we considered elastic scattering of the $n=0$ state with a KK state with nonzero charge. This is a case of unequal mass scattering. 
We adopted the same prescription as we developed for elastic scattering of $n=0$ states. In the brief subsection 4.2.2 we outlined
the argument that this amplitude would satisfy desired and expected analyticity property in the forward direction. Indeed, fixed $t$
dispersion relations can be written down as well. 
 This case has not been studied by Khuri. In the study of the  two cases analyzed in Section 4.2.1 and 4.2.2,   the unitarity
of the S-matrix was not explicitly invoked in the computations. It was sufficient to require Lorentz invariance and microcausality to arrive at those
conclusions. \\
I investigated analyticity property for the forward elastic scattering amplitude 
of states with nonzero KK charges. In principe, one can study elastic scattering of two states carrying KK charges
$(l,n)$. The final particles will also carry the same charge since we focus on elastic process. However, I considered a simpler scenario, without any loss
of generality, when initial (two) particles carry the same KK charge, $n>0$. If these charges were different, say $(l,n)$, then it will be elastic scattering of two
unequal mass particles of respectively masses, $m_l$ and $m_n$. We do not foresee any obstacles to prove forward dispersion relations
in the unequal charge elastic scattering processes. For the equal charge elastic scattering (and hence equal mass elastic scattering) case;  
first we derived an expression for the nonforward scattering amplitude starting
from the LSZ reduction particles with equal  charges .
 Our goal is to derive dispersion relation. Therefore, we extracted the {\it imaginary} part of the amplitude and took the
opportunity to invoke generalized unitarity relation. Thus we derived an expression for product of currents and  we inserted a complete set of states
between the two currents in the expression for the matrix element. The 
complete set of physical states included contributions of the entire KK tower in the matrix element as long as it  respects energy momentum conservation and discrete charge conservation. To 
recollect, the S-matrix is obtained when the four external particles are on the mass shell and all the  intermediate states are physical states. One
important conclusion is that energy momentum conservation (and mass shell condition) ensures that {\it entire KK tower does not contribute}  to
the sum of intermediate states. As long as $s$ is finite, we may choose it to be very large, there is a cut off on the contribution of KK towers. The second
important point to note that we see a glimpse of crossing symmetry (see remarks after (\ref{nn12}))  in the following way: 
The first term in that equation contributes to the $s$-channel process and the second term does not contribute  from energy momentum conservation
consideration. The second term, to be interpreted as {\it crossed channel reaction}, contributed to that reaction and the first term does not.
These conclusions hold good for physical processes. That the two amplitudes coincide in the unphysical kinematical region for real momentum 
variable   is well known.
We have neither proved crossing symmetry here nor we had any intentions to prove it here. Finally, we show that the forward scattering amplitude satisfies
dispersion relations. We invoked arguments of Symanzik in proving the forward dispersion relation.
Our conclusion is that in a relativistic quantum field theory defined on a spacetime
geometry, $R^{3,1}\otimes S^1$, the forward scattering amplitude for elastic scattering scattering of KK states  (with nonzero charge) the dispersion 
relation is satisfied.\\
Now we proceed to discuss some aspects of this investigation. We have assumed existence of stable particles in the entire spectrum of the
the theory defined on $R^{3,1}\otimes S^1$  geometry. Our arguments is based on the conservation of KK discrete charge $q_n={{n}\over R}$; 
it is the momentum along the compatified direction. We have also assumed the 
absence of bound states. The charge, $q_n$  is like a global charge and is assumed to be conserved. 
This conservation law does not originate from a local gauge invariance. At  the 
moment we have no strong argument about absence of bound states. If we had considered a five dimensional theory with gravity and
had utilized KK technique to reduce it to a theory in flat Minkowski space with geometry $R^{3,1}\otimes S^1$, there will a $U(1)$ gauge field coupled
to the resulting KK scalars in the spectrum. The gauge field appears in the standard dimensional reduction of a five dimensional
theory (with gravity)  coupled to a massive neutral scalar field. The spectrum of the flat four dimensional theory is a (zero mode) scalar, the tower 
of KK states and a $U(1)$ gauge field couple to all KK states. In this scenario, the charge could be associate with a $U(1)$ gauge {\it charge}. However, 
we cannot
use LSZ procedure in the {\it five} dimensional curved space. Nonetheless in the compactified theory, defined in a flat  space with an $S^1$ coordinate the 
 LSZ formalism can be incorporated. In this scenario, with one $U(1)$ gauge field, may be we could have bound states with very low binding energy (like BPS states);
although BPS states appear in spontaneously broken symmetry theory (like BPS monopoles).  We mention in passing that  the present investigation
of analyticity of amplitude is carried out for $S^1$ compactification. $S^1$ compactification does not play any special role. We can compactify a
flat space ${\hat D}$ dimensional field theory to $D=4$  theory on a ${\hat D}-4$ dimensional torus \cite{jsjs,jmjhs}. The line of arguments followed in this work
can be suitably generalized to study the analyticity properties of corresponding theory. At this stage we offer no further remarks and  leave the details for future
investigations.\\
Khuri  \cite{khuri2} was motivated by the large extra dimension scenario to undertake the problem.  He had raised the question what will be the consequences
of his conclusions (in the potential scattering model) if indeed the dispersion relation is not valid at LHC energies. However, in the field theory, under the
consideration, 
 the forward dispersion relation holds good. Nevertheless it is important to ask if there is  a large radius compactified theory which
is accessible at LHC energy what are consequences from rigorous field theoretic perspectives?
    
\bigskip
\noindent { \bf Acknowledgements}: It is my great pleasure to thank Andr\'e Martin who inspired me to undertake this investigation. I am grateful to him
for numerous  valuable discussions. I wish to record my thanks to Luis Alvarez Gaume, D. Z. Freedman, Adam Schwimmer and Stefan Theisen for
very illuminating discussions and e-mail correspondances. The gracious hospitality of Hermann Nicolai and the Max-Planck Instit\"ut  f\"ur Gravitational
Physik,  the Albert Einstein Instit\"ut is gratefully acknowledged.

\end{document}